\documentclass[3p,times]{elsarticle}

\usepackage{graphicx}
\usepackage{epsfig}
\usepackage{dcolumn}
\usepackage{bbm}
\usepackage{multirow}
\usepackage[utf8]{inputenc}
\usepackage[colorlinks,linkcolor=red,anchorcolor=green,citecolor=blue,CJKbookmarks=True]{hyperref}
\usepackage[toc]{appendix}

\begin{document}
\begin{frontmatter}
\title{The glueball content of $\eta_c$}

\author[ihep,ucas]{Renqiang Zhang}

\author[ihep]{Wei Sun\corref{cor}}
\ead{sunwei@ihep.ac.cn}

\author[ihep,ucas]{Ying Chen\corref{cor}}
\ead{cheny@ihep.ac.cn}

\author[ihep,ucas]{Ming Gong}

\author[hnnu,ldqsqc,sicqea]{Long-Cheng Gui}

\author[ihep,ucas]{Zhaofeng Liu}

\cortext[cor]{Corresponding author}

\address[ihep]{Institute of High Energy Physics, Chinese Academy of Sciences, Beijing 100049, P.R. China}
\address[ucas]{School of Physics, University of Chinese Academy of Sciences, Beijing 100049, P.R. China}
\address[hnnu]{Department of Physics, Hunan Normal University, Changsha 410081, P.R. China}
\address[ldqsqc]{Key Laboratory of Low-Dimensional Quantum Structures and Quantum Control of Ministry of Education, Changsha 410081, P.R. China}
\address[sicqea]{Synergetic Innovation Center for Quantum Effects and Applications(SICQEA), Hunan Normal University, Changsha 410081, P.R. China}

\begin{abstract}
We carry out the first lattice QCD derivation of the mixing energy and the mixing angle of the pseudoscalar charmonium and glueball on two gauge ensembles with $N_f=2$ degenerate dynamical charm quarks. The mixing energy is determined to be $49(6)$ MeV on the near physical charm ensemble, which seems insensitive to charm quark mass. By the assumption that $X(2370)$ is predominantly a pseudoscalar glueball, the mixing angle is determined to be approximately $4.6(6)^\circ$, which results in a $+3.9(9)$ MeV mass shift of the ground state pseudoscalar charmonium. In the mean time, the mixing can raise the total width of the pseudoscalar charmonium by 7.2(8) MeV, which explains to some extent the relative large total width of the $\eta_c$ meson. As a result, the branching fraction of $\eta_c\to \gamma\gamma$ can be understood in this $c\bar{c}$-glueball mixing framework. On the other hand, the possible discrepancy of the theoretical predictions and the experimental results of the partial width of $J/\psi\to\gamma\eta_c$ cannot be alleviated by the $c\bar{c}$-glueball mixing picture yet, which demands future precise experimental measurements of this partial width. 
\end{abstract}

\end{frontmatter}

\section{Introduction}
The $\eta_c$ meson is usually assigned to be the $1^1S_0$ state of charmonium 
in the quark model. The latest review of the Particle Data Group (PDG)~\cite{Zyla:2020zbs} gives its mass and the total width to be $m_{\eta_c}=2983.9\pm 0.4$ MeV and $\Gamma_{\eta_c}=32.0\pm 0.7$ MeV, respectively. Its width seems quite large among the charmonium states below the $D\bar{D}$ threshold, since its strong decays take place only through the Okubo-Zweig-Iizuka rule (OZI rule)~\cite{Okubo:1963fa,Mandula:1970wz,tHooft:1976rip} suppressed processes. This large width motivates the scenario that $\eta_c$ may have a sizable glueball component. Among the established flavor singlet pseudoscalar mesons, $\eta(1405)$ is usually taken as a candidate for the pseudoscalar glueball~\cite{Cheng:2008ss,He:2009sb,Li:2009rk}. However,
the quenched lattice QCD studies~\cite{Morningstar:1997ff,Morningstar:1999rf,Chen:2005mg,Chowdhury:2014mra} predict that the mass of the pseudsoscalar glueball is around 2.4-2.6 GeV, which is confirmed by lattice simulations with dynamical quarks~\cite{Richards:2010ck,Gregory:2012hu,Sun:2017ipk} (Note that in these lattice studies with dynamical quarks, only gluonic operators built from Wilson loops were used and the possible mixing of glueballs with $q\bar{q}$ mesons and multiple meson states were not considered yet). This raised a question on $\eta(1405)$ as a glueball candidate because of its much lighter mass. On the other hand, there is also a theoretical analysis claiming that $\eta(1405)$ and $\eta(1475)$ can be the same state belong to the $q\bar{q}$-nonet in the 1.3-1.5 GeV mass region~\cite{Wu:2011yx}, such that there is no need of a pseudoscalar glueball state in this region. Given the mass of the pseudoscalar glueball predicted by lattice QCD, it is intriguing to study the possible mixing between the pseudoscalar charmonium and the glueball. Apart from the total width of $\eta_c$, this mixing scenario is also physically relevant to the understanding of $\eta_c$ properties in the $\eta_c\to\gamma\gamma$~\cite{Ackleh:1991dy,Munz:1996hb,Gupta:1996ak,Huang:1996sw} and $J/\psi\to\gamma \eta_c$ processes~\cite{Dudek:2006ej,Dudek:2009kk,Chen:2011kpa,Becirevic:2012dc,Donald:2012ga,Gui:2019dtm}, where there exist more or less tensions between the experimental observations and the theoretical expectations. The phenomenological and lattice studies on this topic can be found in Refs.~\cite{Tsai:2011dp,Qin:2017qes,Eshraim:2018jkt,CLQCD:2016ugl}.

In this work, we investigate the charmonium-glueball mixing relevant to $\eta_c$ in the lattice QCD formalism. There have been pioneering lattice studies of the mixing of the scalar glueball and $q\bar{q}$ mesons~\cite{Lee:1999kv,McNeile:2000xx}. Strictly speaking, this kind of study should be carried out by the lattice calculation with dynamical quarks. It is known that glueballs are flavor singlets and can mix only with flavor singlet $q\bar{q}$ mesons or meson systems. The propagator of a flavor singlet $q\bar{q}$ meson has contributions from both connected and disconnected quark diagrams. Theoretically, in full QCD the disconnected diagrams are the valence quark loops sandwiched with a tower of sea quark loops. When dynamical quarks are absent, the propagator of a flavor singlet $q\bar{q}$ state is not complete such that there are no propagating modes. Since this work is an exploratory study on the charmonium-glueball mixing, we generate the gauge configurations with charm sea quarks and omit light sea quarks for the theoretical simplicity. Our lattice setup is unitary for charm quarks and permits the charmonium-glueball mixing to take place at any time in a temporal interval, since the species of the sea quarks and the valence quarks are the same. In practice, we generate two large gauge ensembles with two degenerate flavors of charm sea quarks. The Large statistics are mandatory for glueballs to have good signal-to-noise ratios. The key task of this study is the calculation of the annihilation diagrams of charm quarks, which is highly computational demanding. For this we adopt the distillation method~\cite{Peardon:2009gh} which enables us to realize the gauge covariant smearing of quark fields and the all-to-all quark propagators simultaneously.  


\section{Numerical Details}\label{sec:numerical}
\subsection{Lattice Setup}
As an exploratory study, we ignore the effect of light quarks and generate gauge configurations 
with $N_f=2$ flavors degenerate charm sea quarks on an $L^3\times T=16^3\times 128$ anisotropic lattice with the aspect ratio being set to $\xi=a_s/a_t=5$, where $a_t$ and $a_s$ are the temporal and spatial lattice spacing, respectively. 
The lattice spacing $a_s$ is determined to be $a_s=0.1026$ fm through the static potential and $r_0=0.491$ fm.
We use the tadpole improved anisotropic clover fermion action and tadpole improved gauge action,
the details of lattice action can be found in~\cite{Sun:2017ipk}.
To investigate the mass dependence of the mixing, we generate two gauge ensembles (denoted by  Ensemble \texttt{I} and Ensemble \texttt{II}) with different bare charm quark masses.
The parameters of the gauge ensembles are listed in Table~\ref{table:config}, where $m_{J/\psi}$ is the corresponding mass of the vector charmonium on these two ensembles. The charm quark mass on Ensemble \texttt{II} is close to the physical one with $m_{J/\psi}=3068$ MeV, which is not far from the experimental $J/\psi$ mass $3097$ MeV. The quark mass on Ensemble \texttt{I} is a little lighter than the physical charm quark mass. We would like to use these two ensembles to check quark mass dependence of our results. In order to get good signals of glueballs, we generate the gauge ensembles with high statistics in this study. 

\begin{table}[t]
	\renewcommand\arraystretch{1.5}
	\small
	\caption{Parameters of two $N_f=2$ gauge ensembles with degenerate charm sea quarks.}
	\label{table:config}
	\begin{center}
		\begin{tabular}{cllllcc}
			\hline
			\texttt{Ensemble} 	& $L^3\times T$       &		$\beta$	&	$a_s$(fm)	&	$\xi$ 	& $N_\mathrm{cfg}$    &	$m_{J/\psi}$(MeV)  	\\\hline
			\texttt{I}		& $16^3 \times 128$   &   	2.8	&	$0.1026$	&	5	            & $ 7000$  &	$2743(1)$		  \\
			\texttt{II}		& $16^3 \times 128$   &		2.8     &	$0.1026$	&	5	        & $ 6084$  &	$3068(1)$		\\
			\hline
		\end{tabular}
	\end{center}
\end{table}

\subsection{Operators and correlation functions}
The principal goal of this work is to investigate the possible mixing of the pseudoscalar glueball and the pseudoscalar $c\bar{c}$ meson, therefore the annihilation diagrams of charm quark and antiquark should be taken care of. For this to be done, we adopt the distillation method~\cite{Peardon:2009gh}:
First, for each configuration and on each time slice, we calculate $N=50$ eigenvectors $v_n(t)$ of the gauge covariant spatial lattice Laplacian operator 
$-\nabla_{xy}^2(t)$. Thus the smeared charm quark field $c^{(s)}$ can be obtained by $c^{(s)}(\mathbf{x},t)=[V(t)V^\dagger(t)]_{\mathbf{xy}} c(\mathbf{y},t)$ where $V(t)$ is a matrix with each column being an eigenvector $v_n(t)$. Secondly, the all-to-all propagator $S^{(c)}_{\alpha\beta}(\mathbf{x},t;\mathbf{y},t')$ of $c^{(s)}$ can be derived through perambulators defined in the framework of the distillation method (one can refer to Ref.~\cite{Zhang:2021xrs} for the technical details).

Physically, there is only one flavor of the charm quark, while we have two degenerate flavors of charm quarks in the fermion action of our setup, therefore two degenerate charm sea quarks in the gauge configuration, which can be denoted by $c_1(x)$ and $c_2(x)$, and compose an `isospin' doublet similar to $u$ and $d$ quarks. Since glueballs are independent of quark flavors and can mix only with flavor singlet mesons, the pseudoscalar charmonium of interest in this work is only the flavor (isospin) singlet state, whose interpolation field can be defined in terms of the smeared charm quark fields $c_1^{(s)}$ and $c_2^{(s)}$ 
\begin{equation}
	\mathcal{O}_\Gamma = \frac{1}{\sqrt{2}}(\bar{c}_1^{(s)}\Gamma c_1^{(s)} + \bar{c}_2^{(s)}\Gamma c_2^{(s)}),
\end{equation}
where $\Gamma$ refers to $\gamma_5$ or $\gamma_5\gamma_4$. Based on the degeneracy of the two flavors of charm quarks, the correlation function of $\mathcal{O}_\Gamma$ can be expressed as 
\begin{eqnarray}\label{eq:ccc}
C_{CC}(t)&=&\frac{1}{T}\sum\limits_{t_s=1}^{T}\sum\limits_{\mathbf{xy}}\langle \mathcal{O}_\Gamma (\mathbf{x},t+t_s)\mathcal{O}_\Gamma ^\dagger(\mathbf{y},t_s)\rangle\nonumber\\
&=&\frac{1}{T}\sum\limits_{t_s=1}^{T}\sum\limits_\mathbf{xy}\langle C(\mathbf{x},t+t_s;\mathbf{y},t_s)+2D(\mathbf{x},t+t_s;\mathbf{y},t_s)\rangle\equiv C(t)+2D(t)
\end{eqnarray} 	
with $C(\mathbf{x},t;\mathbf{y},t')$ and $D(\mathbf{x},t;\mathbf{y},t')$ being the contributions from the connected and disconnected diagrams, respectively,
\begin{eqnarray}\label{eq:c-and-d}
C(\mathbf{x};t,\mathbf{y},t')&=&- \mathrm{Tr}[\Gamma S^{(c)}(\mathbf{x},t;\mathbf{y},t')\Gamma^\dagger S^{(c)}(\mathbf{y},t';\mathbf{x},t)]\nonumber\\
D(\mathbf{x};t,\mathbf{y},t')&=& \mathrm{Tr}[\Gamma S^{(c)}(\mathbf{x},t;\mathbf{x},t)]
\mathrm{Tr}[\Gamma^\dagger S^{(c)}(\mathbf{y},t';\mathbf{y},t')].
\end{eqnarray}  

For the pseudoscalar glueball operator, we adopt the treatment in Ref.~\cite{Morningstar:1999rf,Chen:2005mg} to get the optimized hermitian operator $\mathcal{O}_G(t)=\mathcal{O}^\dagger_G(t)$ coupling mainly to the ground state glueball based 
on different prototypes of Wilson loops and gauge link smearing schemes, Appendix~\ref{appendix:oper} shows details of the operator construction. Thus we have the 
following correlation functions
\begin{eqnarray}\label{eq:corrs}
C_{GG}(t)&=&\frac{1}{T}\sum\limits_{t_s=1}^{T}\langle \mathcal{O}_G(t+t_s)\mathcal{O}_G(t_s)\rangle\nonumber\\
C_{GC}(t)&=&\frac{1}{T}\sum\limits_{t_s=1}^{T}\sum\limits_{\mathbf{x}}\langle \mathcal{O}_G(t+t_s)\mathcal{O}_\Gamma^\dagger(\mathbf{x},t_s)\rangle\nonumber\\
&=& -\frac{\sqrt{2}}{T}\sum\limits_{t_s=1}^{T}\sum\limits_{\mathbf{x}}\langle \mathcal{O}_G(t+t_s)\mathrm{Tr}[\Gamma^\dagger S^{(c)}(\mathbf{x},t_s;\mathbf{x},t_s)]\rangle\nonumber\\
C_{CG}(t)&=&\frac{1}{T}\sum\limits_{t_s=1}^{T}\sum\limits_{\mathbf{x}}\langle \mathcal{O}_\Gamma(\mathbf{x},t+t_s)\mathcal{O}_G(t_s)\rangle=\mp C_{GC}(t)
\end{eqnarray}
where the $\mp$ sign comes from the hermiticity of $\mathcal{O}_{\Gamma}$ and takes the minus sign for $\Gamma=\gamma_5$ (anti-hermitian) and positive sign for $\gamma_5\gamma_4$ (hermitian). 

\section{Mixing angles}\label{sec:formalism}
Strictly speaking, the hadronic states in lattice QCD are the eigenstates $|n\rangle$ of the lattice Hamiltonian $\hat{H}$, which are defined as $\hat{H}|n\rangle=E_n|n\rangle$. For a given quantum number, $|n\rangle$'s span a orthogonal and complete set, namely
$\sum\limits_n |n\rangle\langle n|=1$ with the normalization condition $\langle m|n\rangle=\delta_{mn}$. Therefore, the correlation function $C_{XY}(t)$ of operator $\mathcal{O}_X$ and $\mathcal{O}_Y$ can be parameterized as 
\begin{equation}\label{eq:gc}
C_{XY}(t)\approx \sum\limits_{n\neq 0} \left[ \langle 0|\mathcal{O}_X|n\rangle \langle n|\mathcal{O}^\dagger_{Y}|0\rangle \left(e^{-E_nt}\pm e^{-E_n(T-t)}\right)\right]
\end{equation}
where the $\pm$ sign is for the same and opposite hermiticities of $\mathcal{O}_X$ and $\mathcal{O}_Y$, respectively. 

Since our lattice formalism is unitary for charm quarks, namely, the species of sea quarks and valence quarks are the same, we can choose another complete state set $\{|\alpha_i\rangle, i=1,2,\cdots\}$ as the state basis, such that an eigenstate $|n\rangle $ of $\hat{H}$ can be expressed in terms of $|\alpha_i\rangle$ as 
$|n\rangle =\sum\limits_i C_{ni}|\alpha_i\rangle$
with $\sum\limits_i |C_{ni}|^2=1$. In this sense, one can say that $|n\rangle$ is an admixture of states $|\alpha_i\rangle$ whose fractions are $|C_{ni}|^2$. For the case of this work, we choose the state set $\{|\alpha_i\rangle,i=1,2,\ldots\}$ of flavor singlet pseudoscalars to be 
$|\alpha_i\rangle=|G_1\rangle, |(c\bar{c})_1\rangle, |G_2\rangle, |(c\bar{c})_2\rangle, \ldots$,
where $|G_i\rangle$ and $|(c\bar{c})_i\rangle$ are the $i$-th pure gauge glueball state and the pure $c\bar{c}$ state, respectively. This might be physically meaningful since glueball states are well defined and turn out to exist in the quenched approximation, as well as that charmonium states are usually considered as $c\bar{c}$ bound states in the phenomenological studies. It should be emphasized that this assumption is the prerequisite of the following discussion, and is the common ansatz in the phenomenological mixing models. 

Obviously, the mixing takes place only between glueball states and $c\bar{c}$ states in the state set $\{|\alpha_i\rangle,i=1,2,\ldots\}$. If the dynamics of the mixing can be treated as perturbations, then to the lowest order of the perturbation theory, it may be assumed that the mixing is dominated by that between the nearest glueball state and $c\bar{c}$ state. Thus the Hamiltonian $\hat{H}$ can be expressed as 
\begin{equation}\label{eq:mix_matrix}
\hat{H}=\left(\begin{array}{cc}
m_{G_1} & x_1\\
x_1  & m_{(c\bar{c})_1}\\
\end{array}\right)\oplus \left(\begin{array}{cc}
m_{G_2} & x_2\\
x_2  & m_{(c\bar{c})_2}\\
\end{array}\right)\oplus \cdots
\end{equation}
where $m_{G_i}$ and $m_{(c\bar{c})_i}$ are the masses of the state $|G_i\rangle$ and $|(c\bar{c})_i\rangle$, respectively. The off-diagonal matrix elements $x_i$ of $\hat{H}$ are called {\it mixing energies} which are exactly the transition amplitude between state $|G_i\rangle$ and $|(c\bar{c})_i\rangle$. Thus the eigenstates $|n\rangle=|g_1\rangle,|\eta_1\rangle, |g_2\rangle, |\eta_2\rangle, \ldots$ are related to $|G_i\rangle$ and $|(c\bar{c})_i\rangle$ by 
\begin{equation}\label{eq:mix}
\left(\begin{array}{c}
|g_i\rangle\\
|\eta_i\rangle\\
\end{array}\right)=\left(\begin{array}{cc}
\cos\theta_i & -\sin\theta_i\\
\sin\theta_i & \cos\theta_i\\
\end{array}\right)\left(\begin{array}{c}
|G_i\rangle\\
|(c\bar{c})_i\rangle\\
\end{array}\right)
\end{equation}
where the mixing angles $\theta_i$ have been introduced. The eigenvalues of $\hat{H}$, namely, the masses of $|\eta_i\rangle$ and $|g_i\rangle$ states can be easily derived as $m_{\eta_i} = \bar{m}_i + \Delta_i\delta_i/2$ and $m_{g_i} = \bar{m}_i - \Delta_i\delta_i/2$, where $\bar{m}_i = \frac{1}{2}(m_{G_i} + m_{(c\bar{c})_i})$, $\Delta_i=m_{(c\bar{c})_i}-m_{G_i}$, $\delta_i=\sqrt{1+4x_i^2/\Delta_i^2}$. Accordingly, the mixing angles $\theta_i$ and mass shifts of $|(c\bar{c})_i\rangle$ can be derived as 
\begin{eqnarray}\label{eq:mix-angle}
\sin\theta_i &=& \mathrm{sgn}({x_i}{\Delta_i})\sqrt{\frac{\delta_i-1}{2\delta_i}} = \frac{x_i}{\Delta_i} + \mathcal{O}\left(\frac{x_i^3}{\Delta_i^3}\right),\nonumber\\
\Delta m_{\eta_i} &=& m_{\eta_i} - m_{(c\bar{c})_i} = -\frac{1}{2}\Delta_i + \frac{1}{2}\Delta_i\delta_i \approx \frac{x_i^2}{\Delta_i},
\end{eqnarray}
where $\mathrm{sgn}({x_i}{\Delta_i})$ refers to the sign of $x_i\Delta_i $.
Therefore, for the ground state $\eta_c(1S)$ we are interested in, the key task is to extract $\theta_1$ that is the mixing angle of the ground state of the pseudoscalar glueball 
$|G_1\rangle$ and the ground state pseudoscalar charmonium $|(c\bar{c})_1\rangle$.
\begin{figure*}[t]
	\centering
	\includegraphics[width=0.3\linewidth]{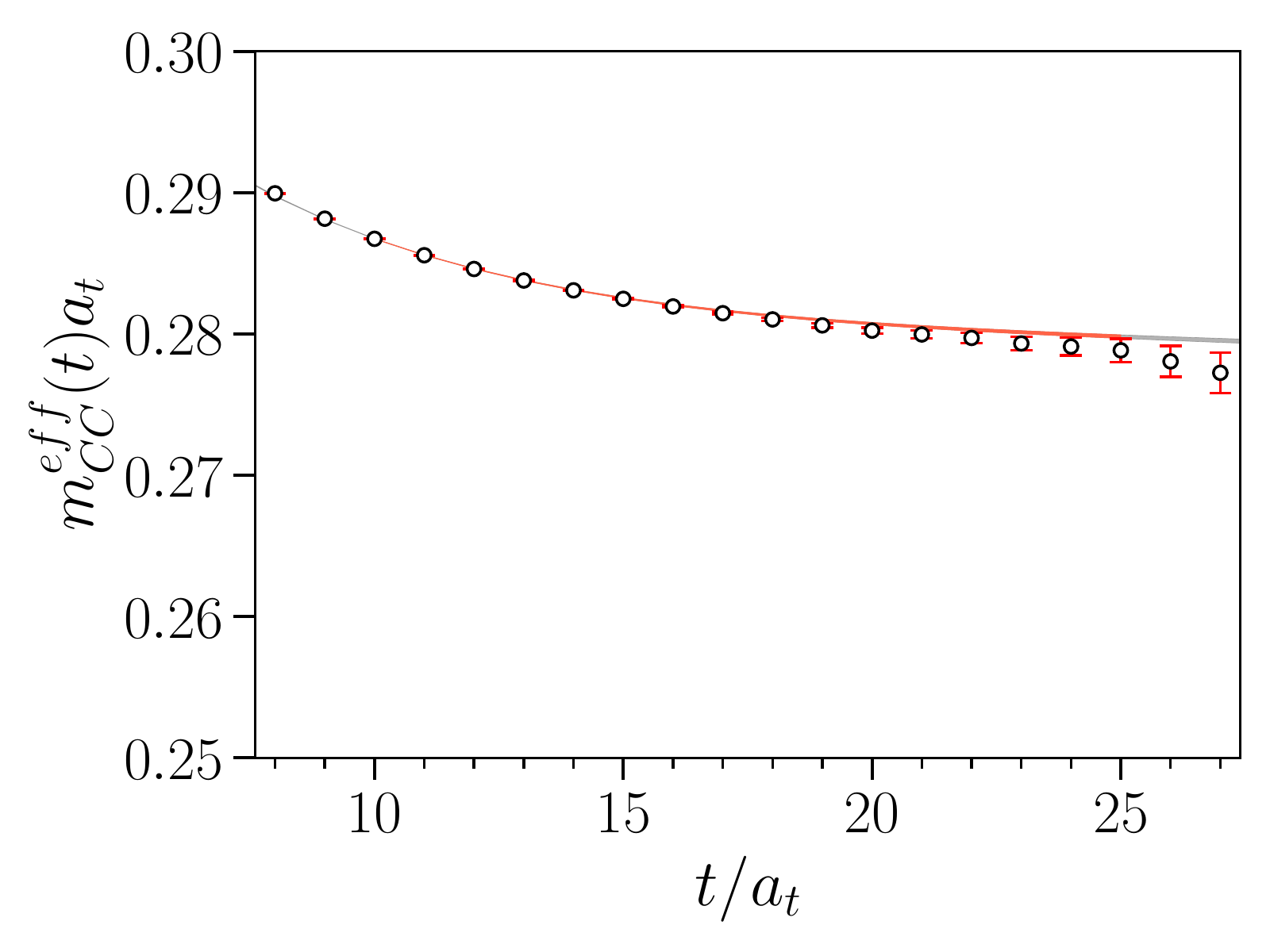}
	\includegraphics[width=0.3\linewidth]{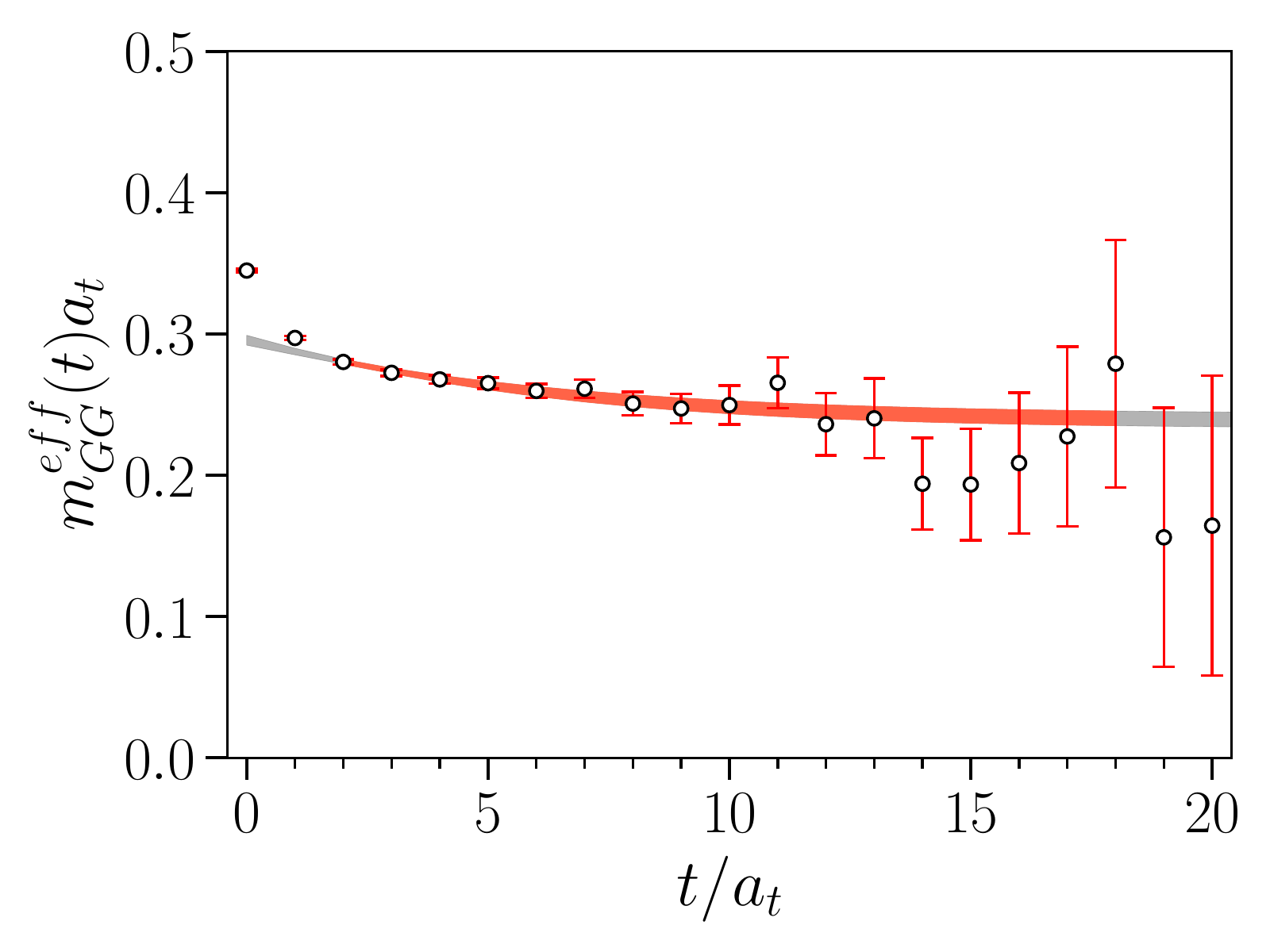}
	\includegraphics[width=0.3\linewidth]{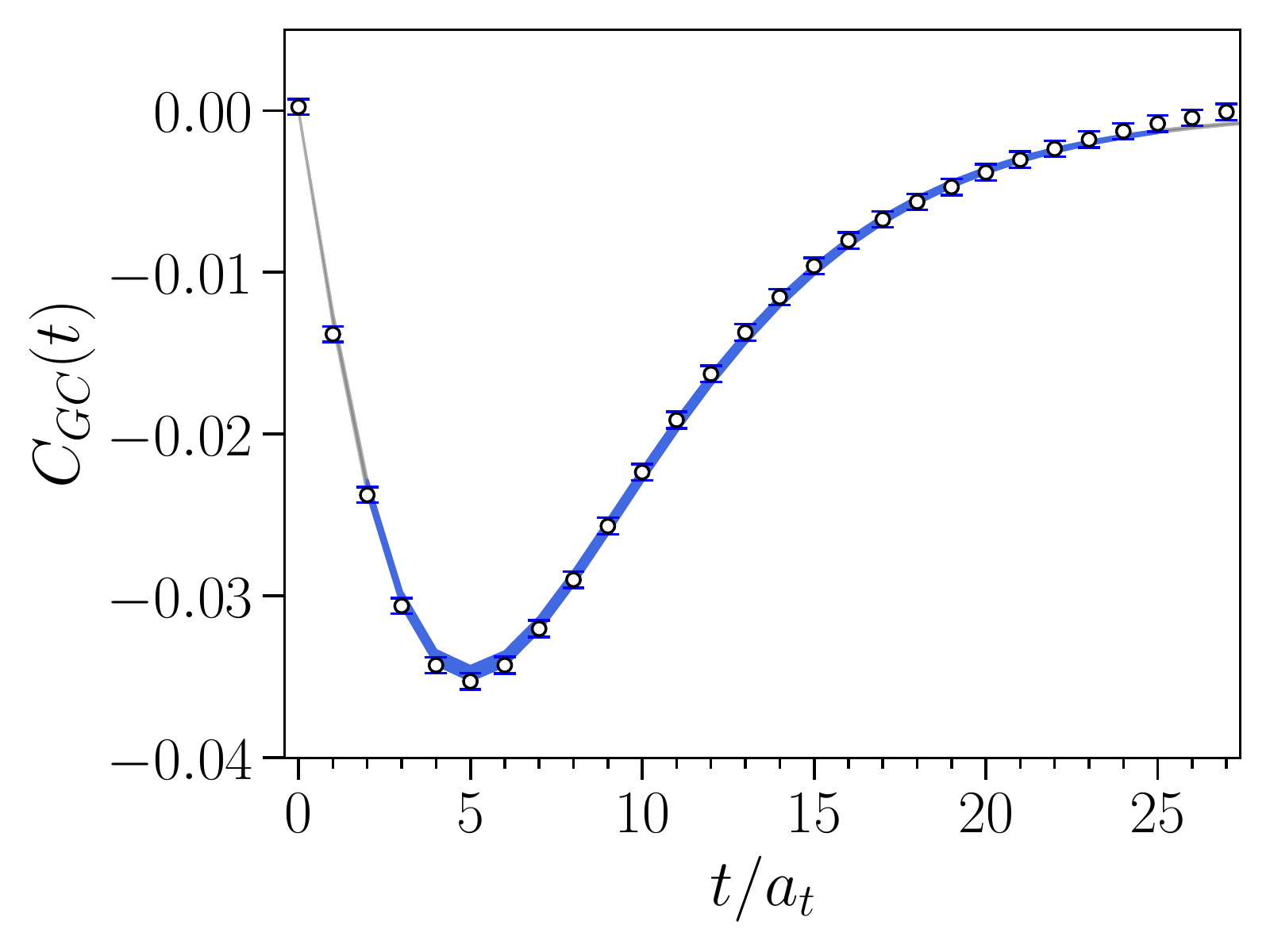} \\
	\includegraphics[width=0.3\linewidth]{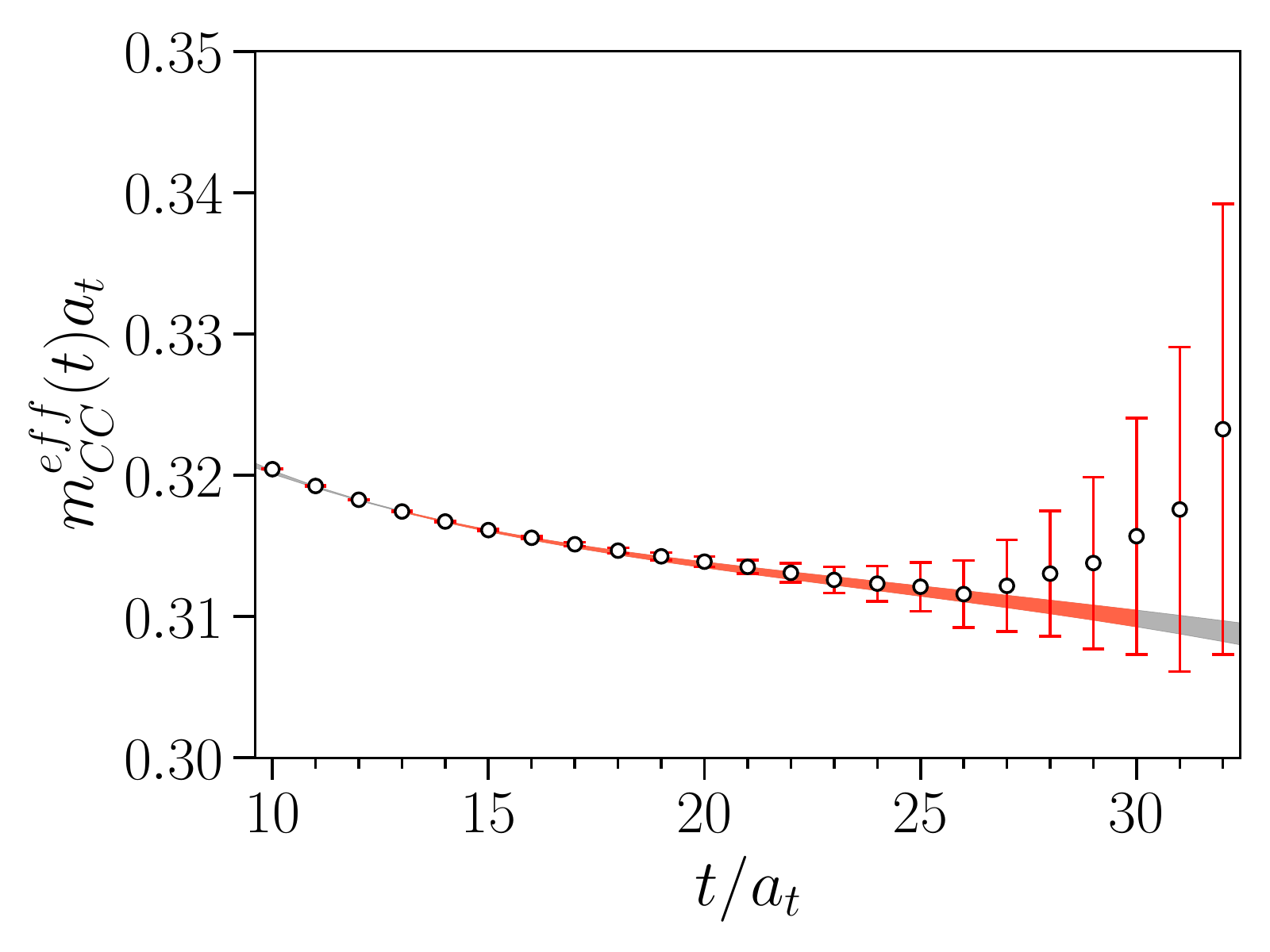}
	\includegraphics[width=0.3\linewidth]{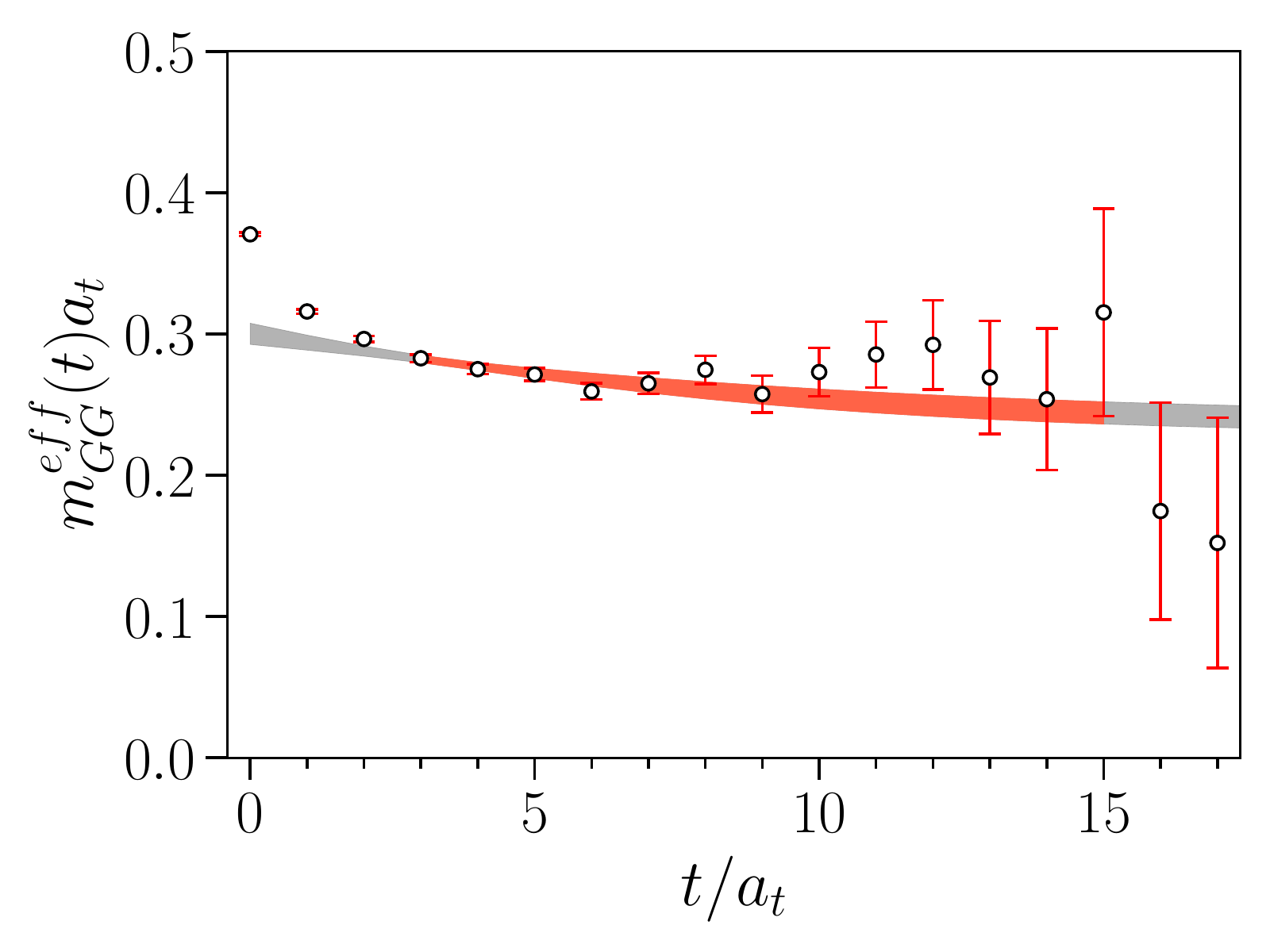}
	\includegraphics[width=0.3\linewidth]{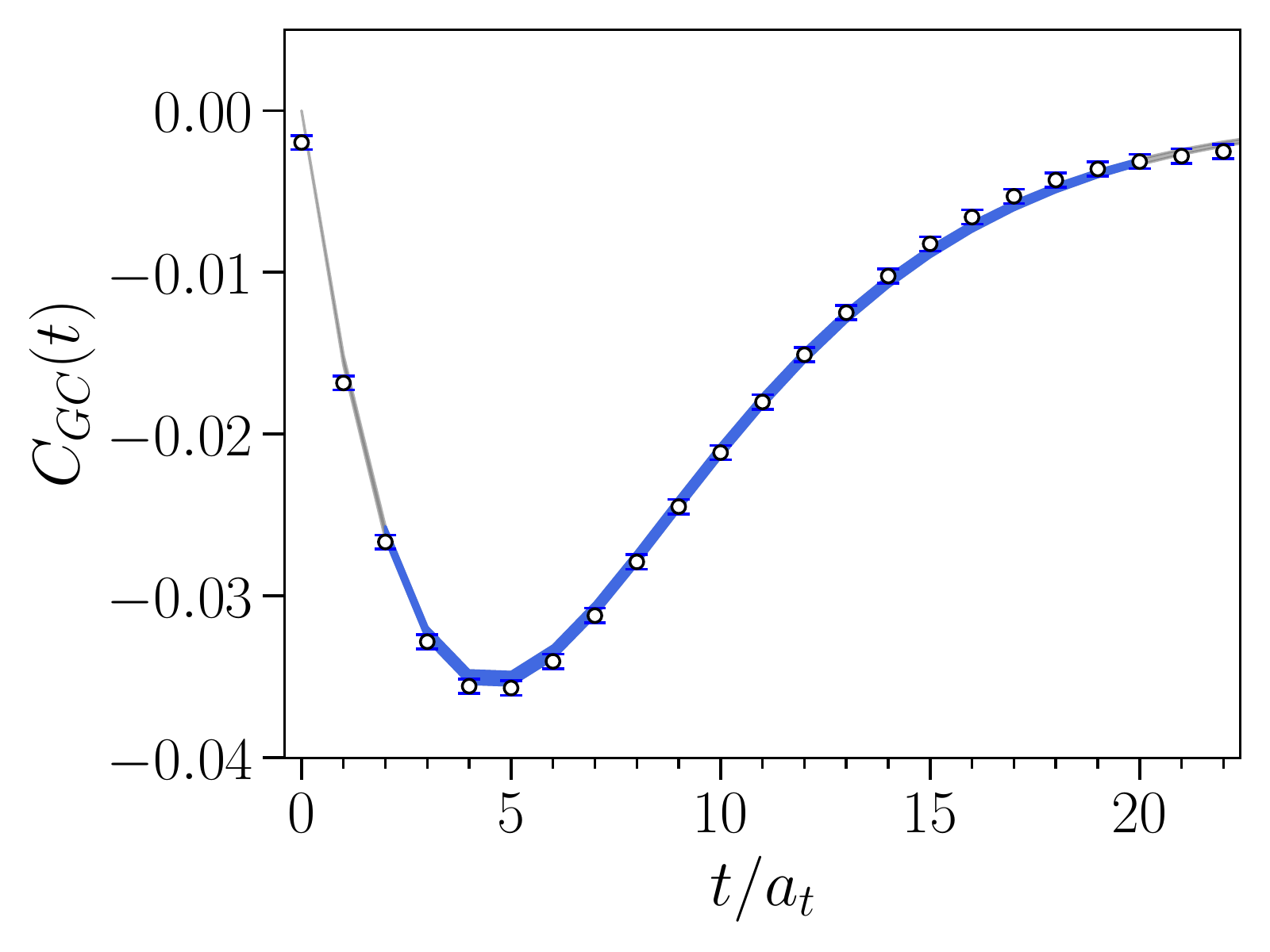}
	\caption{\label{fig:effemgamma5}Effective mass from two point functions 
		$C_{CC}(t)$, $C_{GG}(t)$ and correlation function $C_{GC}(t)$ for operator with $\Gamma=\gamma_5$ using best fit parameters from Eq.~(\ref{eq:gc_general}) and (\ref{eq:gg-gamma5})
		on ensemble \texttt{I} (top) and ensemble \texttt{II} (bottom),
		where points with error bar are from simulation data with jackknife estimated error, the light gray band shows the
		fitted results with best fit parameters in Table.~\ref{table:fit}, and the color band indicates the fitting range. The large errors of $m^\mathrm{eff}_{CC}(t)$ in the time range beyond $t/a_t>20$ come mainly from the disconnected diagrams (see discussions in Appendix~\ref{appendix:comparison}).  }
\end{figure*}
\subsection{The $\Gamma=\gamma_5$ case}
Actually the mixing angle $\theta_i$ can be derived from the correlation function $C_{CG}(t)$ or $C_{GC}(t)$ for the $\Gamma=\gamma_5$ case if we assume boldly that $\mathcal{O}_G$ couples almost exclusively with $|G_i\rangle$ and $\mathcal{O}_{\Gamma}$ couples exclusively with $|(c\bar{c})_i\rangle$, namely,
\begin{eqnarray}\label{eq:create}
\mathcal{O}_G^\dagger|0\rangle&=&\sum\limits_{i\neq 0}\sqrt{Z_{G_i}}|G_i\rangle\nonumber\\
\mathcal{O}_{\gamma_5}^\dagger|0\rangle&=&\sum\limits_{i\neq 0}\sqrt{Z_{(\gamma_5),i}}|(c\bar{c})_i\rangle
\end{eqnarray}
With this assumption and by utilizing Eq.(\ref{eq:mix}) and Eq.(\ref{eq:create}), one has 
\begin{eqnarray}\label{eq:gc_general}
C_{GC}(t)=&&-\sum_i\sqrt{Z_{G_i}Z_{(\gamma_5),i}} \cos\theta_i \sin\theta_i \left(e^{-m_{g_i}t}- e^{-m_{g_i}(T-t)}-(e^{-m_{\eta_i}t}- e^{-m_{\eta_i}(T-t)})\right).
\end{eqnarray}
Note that in the above equation, we also use the fact that $\mathcal{O}_{\gamma_5}$  is anti-hermitian. When $T$ is large, the above parameterization of $C_{GC}(t)$ requires $C_{GC}(t=0)\approx 0$, which is a direct consequence of the assumption of Eq.~(\ref{eq:create}). The measured $C_{GC}(t)$'s from the ensemble I and II are shown in the right most column of Fig.~\ref{fig:effemgamma5}, where one can see that 
this $C_{GC}(t=0)\approx 0$ is meet. This manifests that the assumptions in Eq.~(\ref{eq:gc_general}) are reasonable. 

In order for $\theta_i$'s to be extracted using Eq.~(\ref{eq:gc_general}), one has to know the parameters $m_{g_i}$, $m_{\eta_i}$, $Z_{G_i}$ and $Z_{(\gamma_5),i}$, which, based on the assumptions of Eq.~(\ref{eq:create}), are encoded in 
the correlation functions $C_{CC}(t)$ and $C_{GG}(t)$ as
\begin{eqnarray}\label{eq:gg-gamma5}
C_{GG}(t) &=& \sum\limits_i Z_{G_i}\left[\cos^2\theta_i \left(e^{-m_{g_i}t}+e^{-m_{g_i}(T-t)}\right) + \sin^2\theta_i \left(e^{-m_{\eta_i}t}+e^{-m_{\eta_i}(T-t)}\right) \right]\nonumber\\
C_{CC}(t) &=& \sum\limits_i Z_{(\gamma_5),i}\left[\cos^2\theta_i \left(e^{-m_{\eta_i}t}+e^{-m_{\eta_i}(T-t)}\right)  + \sin^2\theta_i \left(e^{-m_{g_i}t}+e^{-m_{g_i}(T-t)}\right)\right]. 
\end{eqnarray}
Therefore, we carry out a simultaneous fit to $C_{GC}(t)$, $C_{CC}(t)$ and $C_{GG}(t)$ through 
the correlated minimal-$\chi^2$ fitting procedure using the function forms in Eq.~(\ref{eq:gc_general}) and (\ref{eq:gg-gamma5}). Since we focus on $\theta_1$, in practice we only consider the contribution from the lowest two glueball states and two $c\bar{c}$ states, namely, we use $i=1,2$ in above functions to model the data and treat the second states to be the effective states that take account of the contribution of all the higher states. 
\begin{table*}[t]
	\renewcommand\arraystretch{1.5}
	\small
	\caption{Ground state mass and mixing angle fitted from operators with $\Gamma=\gamma_5$ and $\Gamma=\gamma_5\gamma_4$
		on ensemble \texttt{I} and ensemble \texttt{II}, rows started with \texttt{avg.} are the final weighted average results.}
	\label{table:fit}
	\begin{center}
		\begin{tabular}{clcccc|cccc}
			\hline
			\texttt{ensemble} 	&	$\Gamma$	&	$[t_l,t_h]_{CC}$	&	$[t_l,t_h]_{GG}$	&	$[t_l,t_h]_{GC}$	&	$\chi^2/dof$	& $m_{\eta_1}$(MeV)  &	$m_{g_1}$(MeV)	 &	$\theta_1$	& $x_1$(MeV)	\\\hline
			\multirow{3}{*}{\texttt{I}}	&	$\gamma_5$		&	[10, 25]			&	[2, 18]				&	[2, 25]			&	$1.1$	& $2705(2)$   	&  $2289(50)$	 &	$6.8(9)^\circ$  &  49(9)			\\
			&	$\gamma_5\gamma_4$ 	&	[10, 25]			&	[2, 18]				&	[2, 30]				&	$0.98$		& $2701(1)$   	     &  $2283(51)$	 &	$6.5(9)^\circ$		&  48(9)	\\
			\cline{2-10}
			&	\texttt{avg.}	&	---				&	---				&	---				&	---			& $2703(1)$	     &	$2286(50)$	 &	$6.6(9)^\circ$ &  48(9)				\\
			\hline
			\multirow{3}{*}{\texttt{II}}	&	$\gamma_5$		&	[13, 30]			&	[3, 15]				&	[2, 20]		&	$1.1$	& $3028(8)$          &	$2261(74)$	 &	$4.5(6)^\circ$	& 60(10)			\\
			&	$\gamma_5\gamma_4$ 	&	[13, 30]			&	[2, 15]				&	[1, 30]			&	$1.1$	& $3031(3)$   	     &	$2348(47)$	 &	$3.9(3)^\circ$		&  47(5)		\\
			\cline{2-10}
			&	\texttt{avg.}	&	---				&	---				&	---				&	---		& $3031(3)$	     &  $2323(55)$	 &	$4.3(4)^\circ$	 & 49(6)			\\
			\hline
		\end{tabular}
	\end{center}
\end{table*}
The calculated results and fit results are shown in Fig.~\ref{fig:effemgamma5}. The data points in the left most column show the effective masses $m_{CC}^{\mathrm{eff}}(t)$ of the correlation function $C_{CC}(t)$ on the two ensembles (the upper panel is for ensemble I and the lower one is for ensemble II), 
which are defined by 
\begin{equation}
m_{CC}^{\mathrm{eff}}(t)=\ln \frac{C_{CC}(t)}{C_{CC}(t+1)}.
\end{equation}   
The effective mass $m_{GG}^{\mathrm{eff}}(t)$ on the two ensembles are shown as data points in the middle column. The right most two panels of Fig.~\ref{fig:effemgamma5} show the correlation functions $C_{GC}(t)$ obtained on the two ensembles. The curves with error bands are plotted using the best fit parameters obtained through the fitting procedure mentioned above, where the colored bands illustrate the fitting time range. The fitted masses and the mixing angles $\theta_1$ are collected in Table~\ref{table:fit}, where the masses are converted into the values in physical units through the lattice spacings in Table~\ref{table:config}. The fit time windows and the related $\chi^2$ per degree of freedom ($\chi^2/dof$) are also presented. It is seen that the function forms of Eq.~(\ref{eq:gc_general}) and Eq.~(\ref{eq:gg-gamma5}) describes the data very well with a reasonable $\chi^2/dof$. On the ensemble I, the fitted $\eta_c$ mass is around $m_{\eta_1}\approx 2.7$ GeV, while the result on ensemble II is $m_{\eta_1}\approx 3.0$ GeV and close to the experimental value. On the two ensembles, the fitted pseudoscalar glueball mass is around $2.3$ GeV and shows little dependence of the charm quark masses. Finally, we get the mixing angle $\theta_1= 6.8(9)^\circ$ and $\theta_1= 4.5(6)^\circ$ on ensemble I and ensemble II, respectively.
According to Eq.~(\ref{eq:mix-angle}) and using the mass differences $m_{\eta_1} - m_{g_1}$ listed in Table.~\ref{table:fit} as the estimate for $\Delta_1=m_{(c\bar{c})_1}-m_{G_1}$, the mixing energy $x_1$ are derived to be $49(9)$ MeV and $60(10)$ MeV on these two ensembles. 

\begin{figure*}[t]
	\centering
	\includegraphics[width=0.3\linewidth]{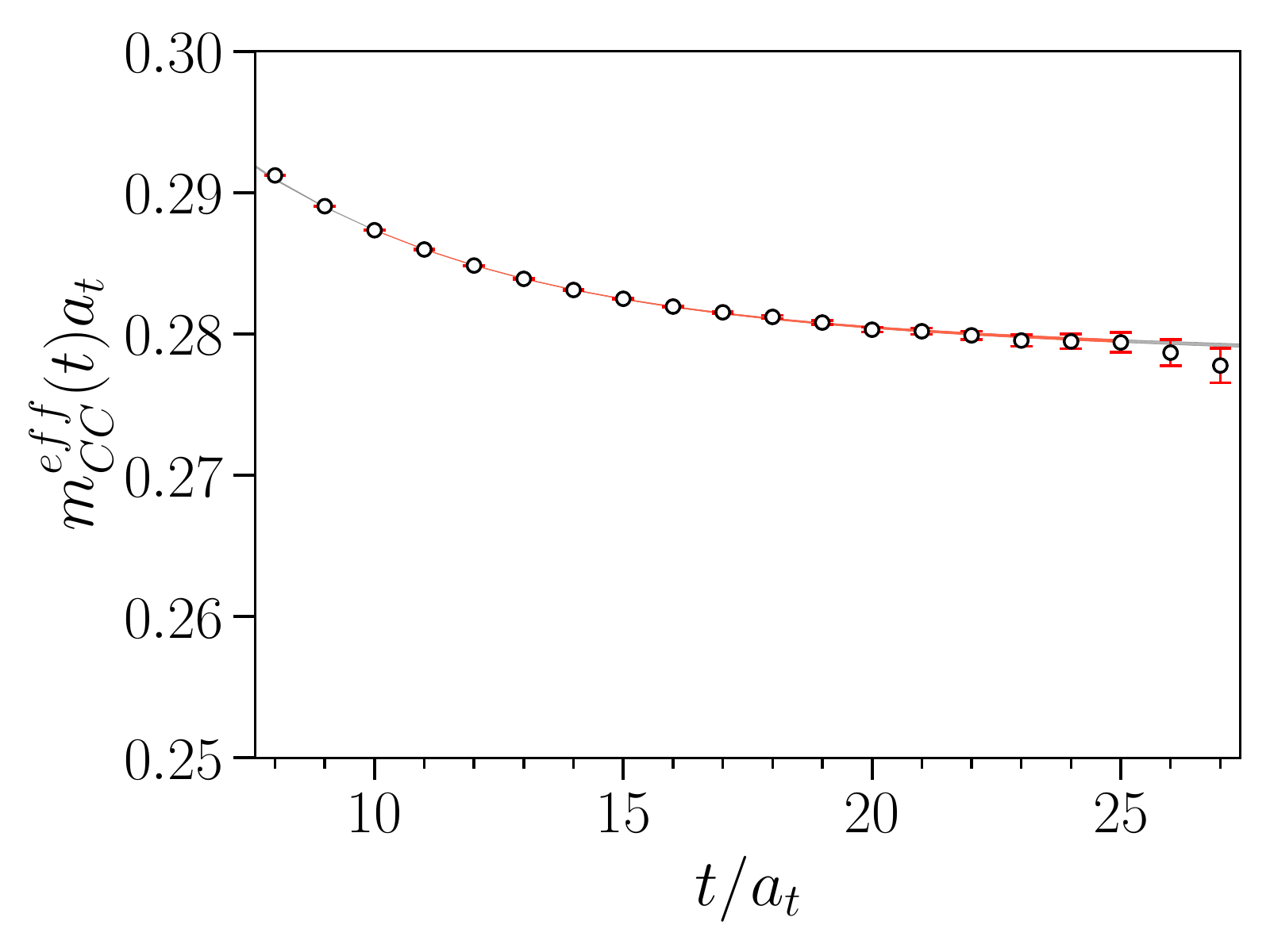}
	\includegraphics[width=0.3\linewidth]{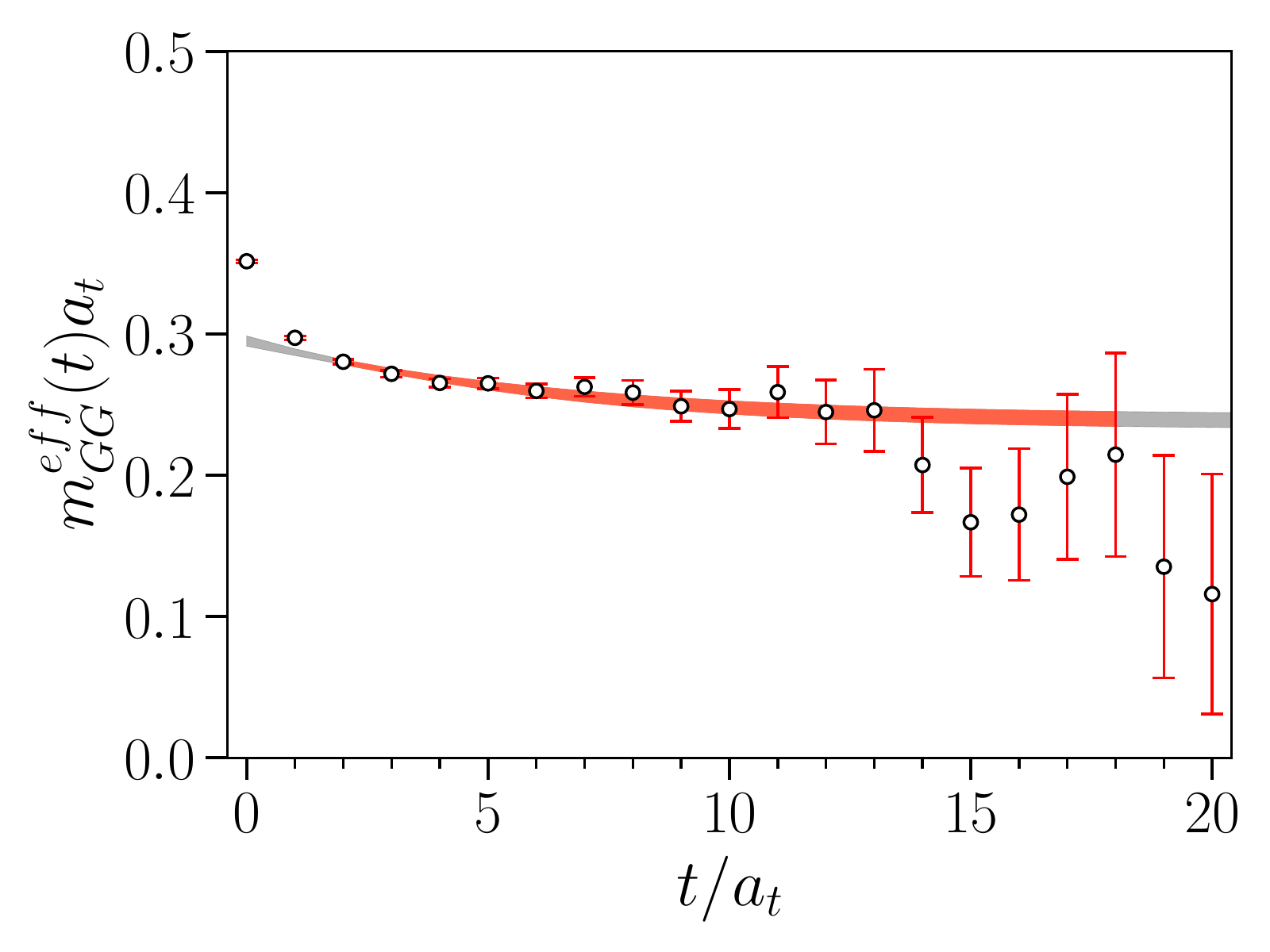}
	\includegraphics[width=0.3\linewidth]{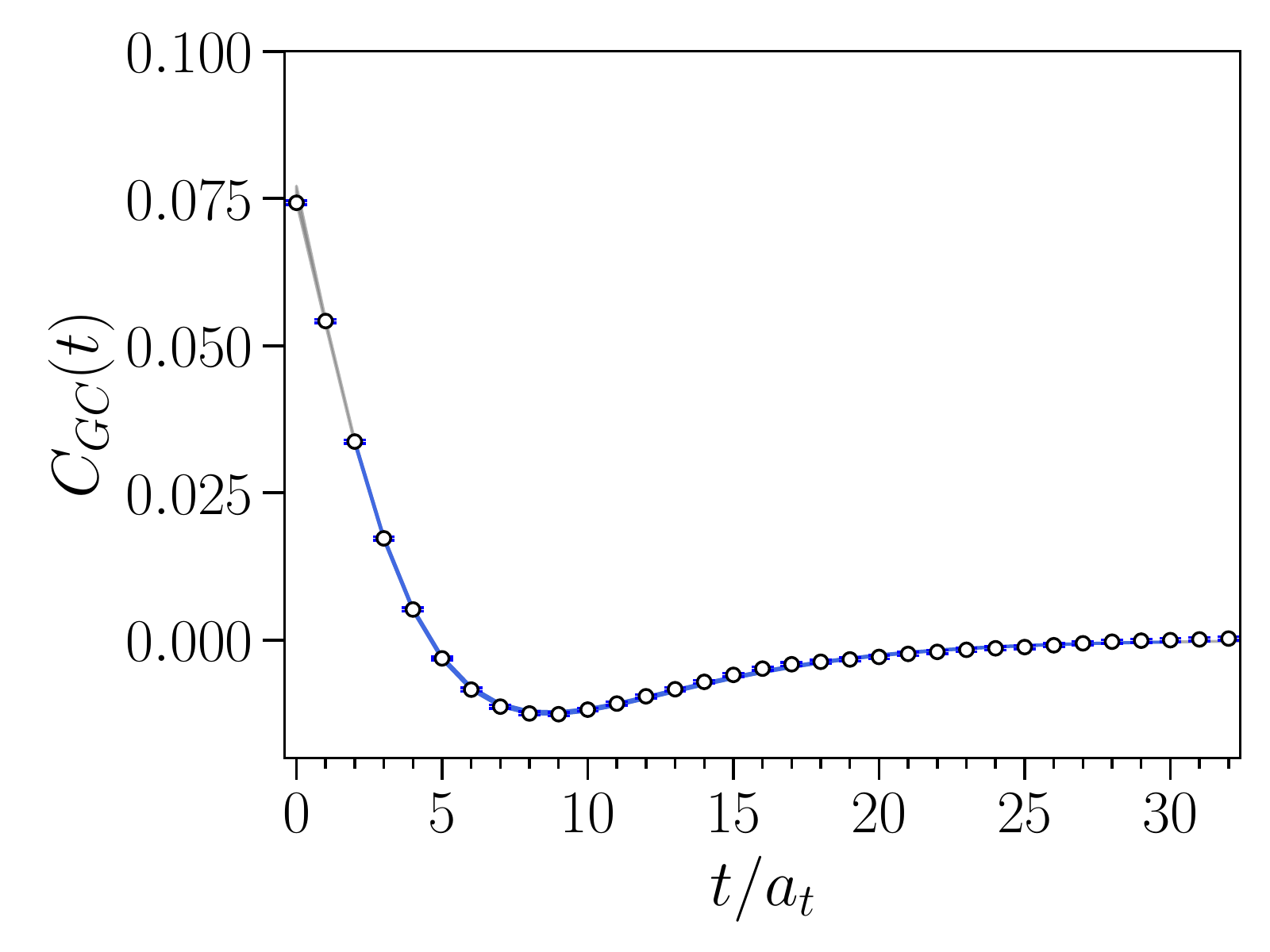} \\
	\includegraphics[width=0.3\linewidth]{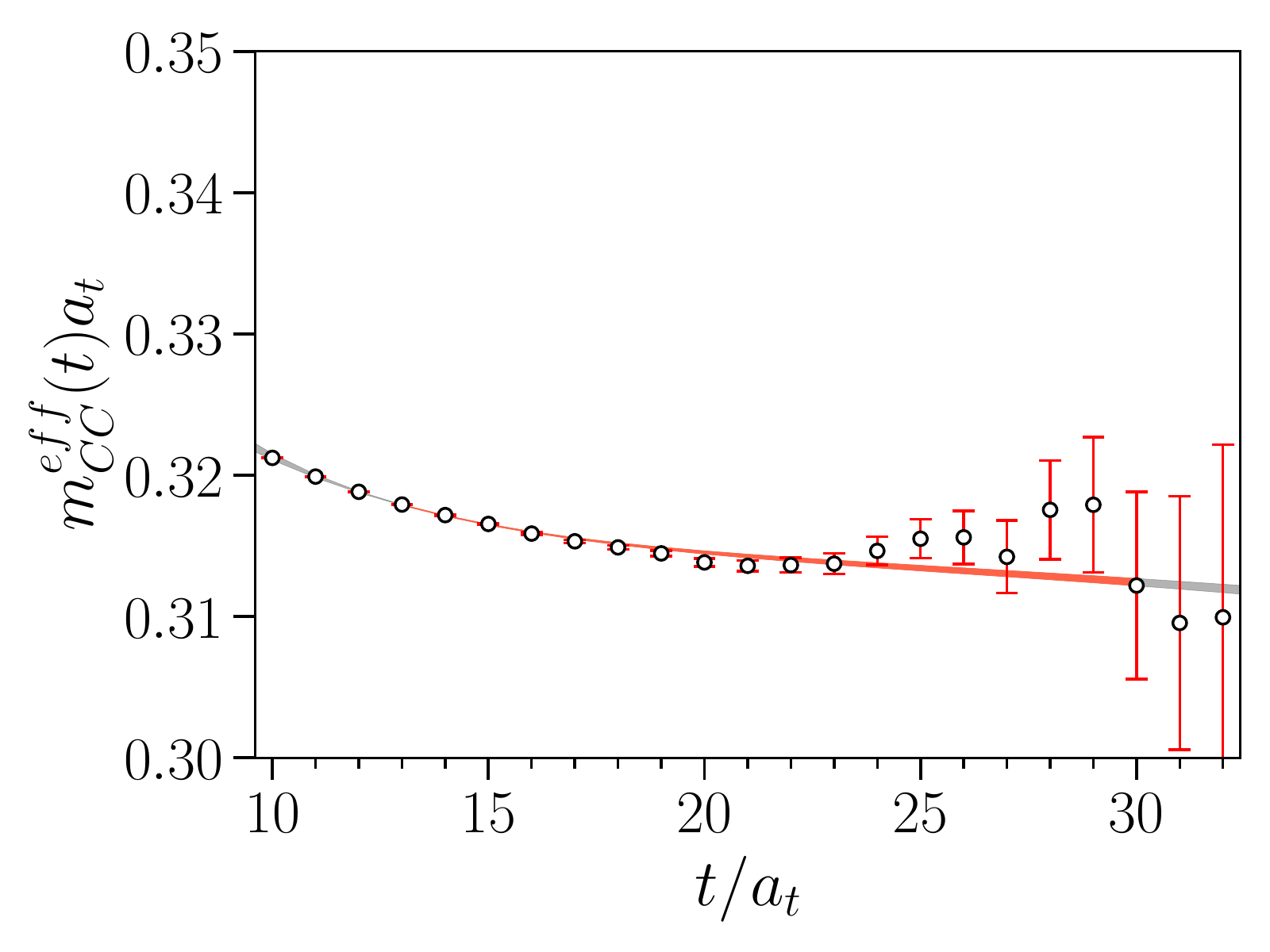}
	\includegraphics[width=0.3\linewidth]{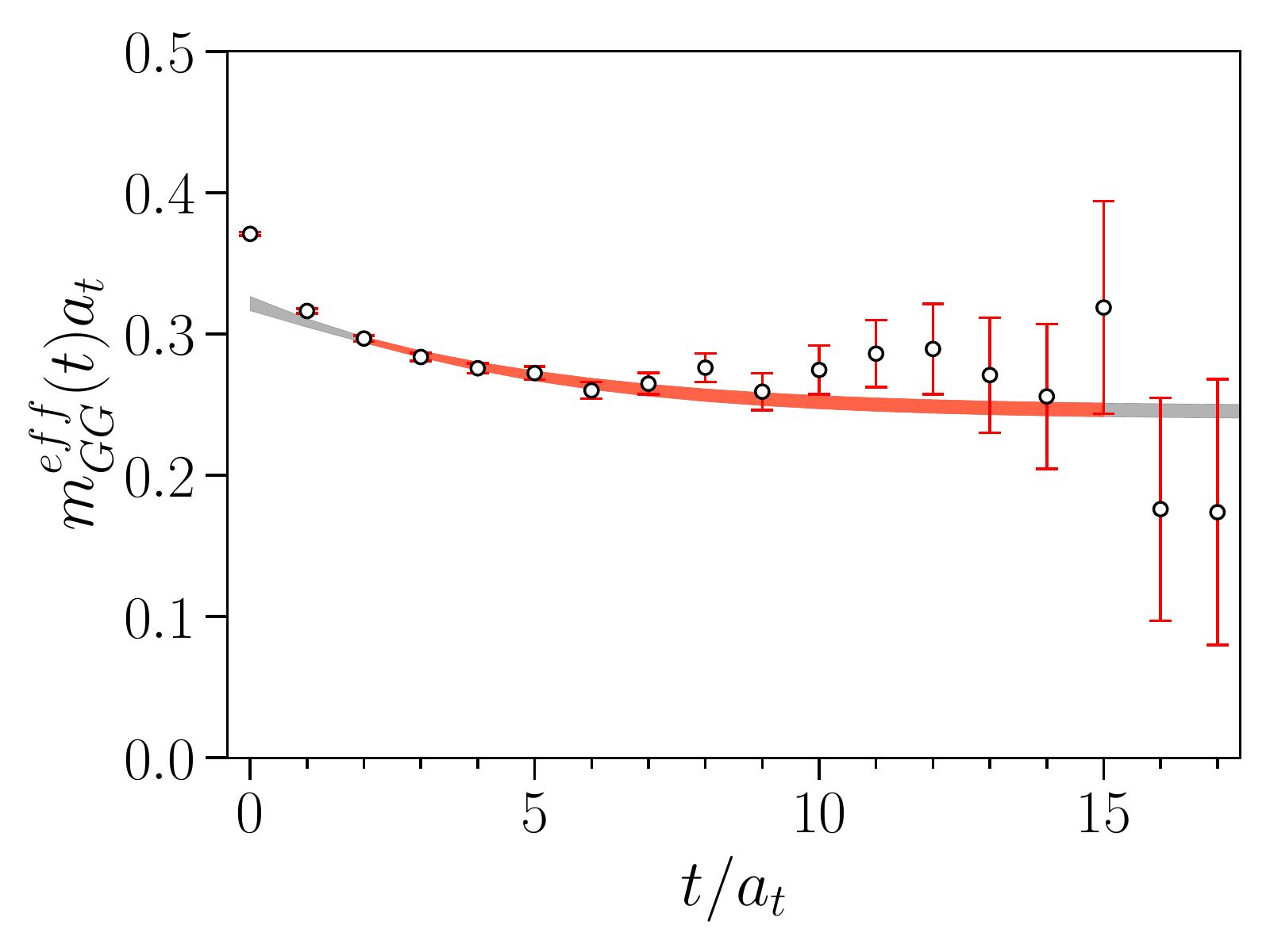}
	\includegraphics[width=0.3\linewidth]{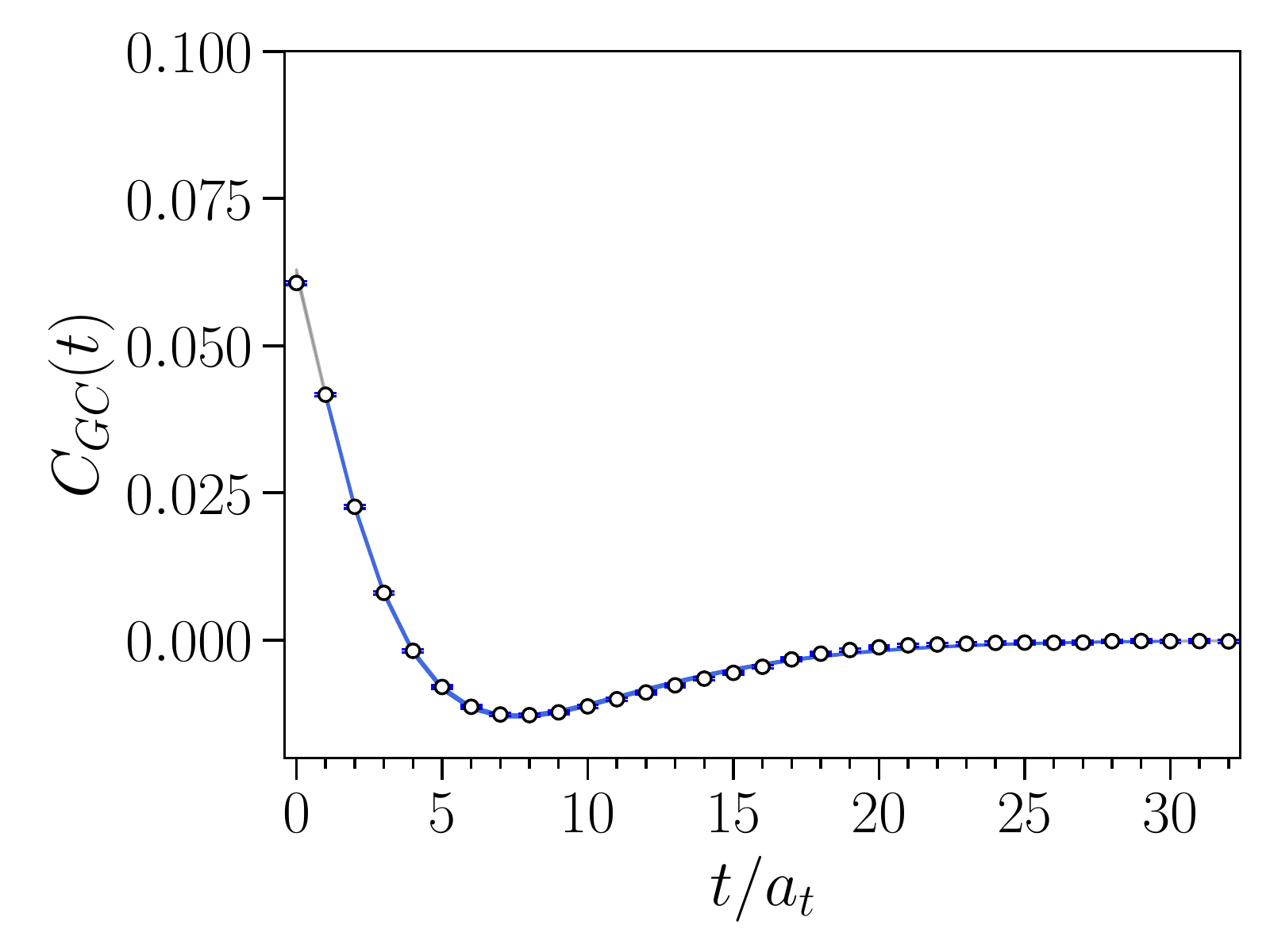}
	\caption{\label{fig:effemgamma45}Effective mass from two point functions 
		$C_{CC}(t)$, $C_{GG}(t)$ and correlation function $C_{GC}(t)$ for operator with $\Gamma=\gamma_5\gamma_4$ 
		using best fit parameters from $C_{GG}(t)$ of Eq.~(\ref{eq:gg-gamma5}) and (\ref{eq:gamma45})
		on ensemble \texttt{I} (top) and ensemble \texttt{II} (bottom),
		where points with error bar are from simulation data with jackknife estimated error, the light gray band shows the
		fitted results with best fit parameters in Table.~\ref{table:fit}, and the color band indicates the fitting range.}
\end{figure*}
\subsection{The $\Gamma=\gamma_5\gamma_4$ case}
As a cross check, we also carried out the similar calculation by using the $\Gamma=\gamma_5\gamma_4$ for the interpolation field operator of the pseudoscalar $c\bar{c}$ states. The corresponding correlation functions $C_{CC}(t)$ and $C_{GC}(t)$ 
are calculated using Eq.~(\ref{eq:corrs}). The effective masses $m_{CC}^{\mathrm{eff}}(t)$ , $m_{GG}^\mathrm{eff}(t)$ and $C_{GC}(t)$ on the two ensembles are shown in Fig.\ref{fig:effemgamma45}. It is interesting to see that, in contrast to the case of $\Gamma=\gamma_5$, the correlation function $C_{GC}(t)$ does not go to zero when $t\to 0$ now (see the right most column of Fig.~\ref{fig:effemgamma45}). This implies that the assumptions in Eq.~(\ref{eq:create}) may not apply here. If we insist the relation $\mathcal{O}_G^\dagger|0\rangle=\sum\limits_{i\neq 0}\sqrt{Z_{G_i}}|G_i\rangle$ still holds, then the second assumption in  Eq.~(\ref{eq:create}) should be modified. 

Actually, the operator $\mathcal{O}_{\gamma_5\gamma_4}$ is the temporal component of the isoscalar axial vector current $J_5^\mu=\bar{c}\gamma_5\gamma^\mu c$ with $c=(c_1,c_2)^T$ here (up to a normalization factor since the charm quark fields in $\mathcal{O}_{\gamma_5\gamma_4}$ are spatially smeared). According to the $U_A(1)$ anomaly of QCD, $J_5^\mu$ satisfies the following anomalous axial vector relation
\begin{equation}\label{eq:u1a}
	\partial_\mu J_5^\mu(x) = 2m_c\bar{c}(x)\gamma_5 c(x) + q(x),
\end{equation}
where $q(x)=\frac{g^2}{16\pi^2} \epsilon^{\alpha\beta\rho\sigma} G_{\alpha\beta}^a G^a_{\rho\sigma}$ is the anomalous term from the $U_A(1)$ anomaly with $g$ being the strong coupling constant and $G_{\alpha\beta}^a$ being the strength of color fields.  The first term on the right hand side of Eq.~(\ref{eq:u1a}) is proportional to our operator $\mathcal{O}_{\gamma_5}$, thus based on the assumption in Eq.~(\ref{eq:create}) we have 
\begin{equation}
\langle 0|\partial_\mu J_5^\mu(x)|G_i\rangle \approx \langle 0|q(x)|G_i\rangle.
\end{equation}
On the other hand, if we introduce the decay constant of the glueball state $|G_i\rangle$ 
through the definition
\begin{equation}\label{eq:fG}
\langle 0|J_5^\mu (x)|G_i,p\rangle=if_{G_i}p^\mu e^{-ip\cdot x},
\end{equation}
then we have 
$\langle 0|\partial_\mu J_5^\mu(0)|G_i,\mathbf{p}=0\rangle=m_{G_i}^2 f_{G_i}$ 
and therefore $f_{G_i}=\frac{1}{m_{G_i}^2}\langle 0|q(0)|G_i\rangle$. Thus from Eq.~(\ref{eq:fG}) we can estimate 
that 
\begin{equation}
\langle 0|\mathcal{O}_{\gamma_5\gamma_4}|G_i,\mathbf{p}=0\rangle \propto \frac{1}{m_{G_i}}\langle 0|q(0)|G_i\rangle.
\end{equation}
Previous lattice studies show that pseudoscalar states can be accessed by the operator $q(x)$~\cite{Chen:2005mg,Chowdhury:2014mra}, thus the nonzero matrix element $\langle 0|q(0)|G_i\rangle$ implies the coupling $\langle 0| \mathcal{O}_{\gamma_5\gamma_4}|G_i\rangle \neq 0$. Consequently we have the following matrix elements
\begin{eqnarray}
\langle 0|\mathcal{O}_{\gamma_5\gamma_4}|g_i\rangle &=& \cos \theta_i \langle 0|\mathcal{O}_{\gamma_5\gamma_4}|G_i\rangle-\sin \theta_i \langle 0|\mathcal{O}_{\gamma_5\gamma_4}|(c\bar{c})_i\rangle\nonumber\\
\langle 0|\mathcal{O}_{\gamma_5\gamma_4}|\eta_i\rangle &=& \sin \theta_i \langle 0|\mathcal{O}_{\gamma_5\gamma_4}|G_i\rangle+\cos \theta_i \langle 0|\mathcal{O}_{\gamma_5\gamma_4}|(c\bar{c})_i\rangle.\nonumber\\
\end{eqnarray}
Thus the correlation function $C_{GC}(t)$ can be parameterized as 
\begin{eqnarray}\label{eq:gamma45}
	C_{GC}(t) &=& \sum_{i=1}^2 \left[\langle 0|\mathcal{O}_G|g_i\rangle \langle g_i|\mathcal{O}_{\gamma_5\gamma_4}|0\rangle \left(e^{-m_{g_i}t} + e^{-m_{g_i}(T-t)}\right) + 
	              \langle 0 |\mathcal{O}_G|\eta_i\rangle \langle \eta_i|\mathcal{O}_{\gamma_5\gamma_4}|0\rangle \left(e^{-m_{\eta_i}t} + e^{-m_{\eta_i}(T-t)}\right)\right] \nonumber\\
	          &\approx & \sqrt{Z_{G_1}}\langle 0|\mathcal{O}_{\gamma_5\gamma_4}|G_1\rangle \cos^2 \theta_1 \left( e^{-m_{g_1}t} + e^{-m_{g_1}(T-t)}\right) \nonumber\\
	          && - \sum_{i=1}^{2}\sqrt{Z_{G_i}Z_{(\gamma_5\gamma_4),i}} \cos\theta_i \sin\theta_i \left(e^{-m_{g_i}t}+e^{-m_{g_i}(T-t)}-(e^{-m_{\eta_i}t}+e^{-m_{\eta_i}(T-t)})\right),
\end{eqnarray}
where $\sqrt{Z_{(\gamma_5\gamma_4),i}}=\langle 0|\mathcal{O}_{\gamma_5\gamma_4}|(c\bar{c})_i\rangle$ has been defined. The second equality is derived based on first assumption of Eq.~(\ref{eq:create}), namely $\langle 0|\mathcal{O}_G|(c\bar{c})_i\rangle=0$,
and $\langle 0| \mathcal{O}_{\gamma_5\gamma_4}|G_i\rangle \neq 0$ discussed above. In the first term of the last expression of Eq.~(\ref{eq:gamma45}), we only keep the mass term of $|g_1\rangle$ state to take care of the temporal behavior of $C_{GC}(t)$ in the early time range. This is justified since the operator $\mathcal{O}_G(t)$ is optimized to couple most to the ground state with $Z_{G_1}\gg Z_{G_2}$. The terms proportional to $\sin^2\theta_i$ are expected to be small and also neglected here.

Using Eq.~(\ref{eq:gg-gamma5}) (after replacing $Z_{(\gamma_5),i}$ by $Z_{(\gamma_5\gamma_4),i}$) and Eq.~(\ref{eq:gamma45}), we carry out a simultaneous fit to $C_{GG}(t)$, $C_{GC}(t)$ and $C_{CC}(t)$. The fit procedure is the same as the case of $\Gamma=\gamma_5$. In Fig.~\ref{fig:effemgamma45}, the colored curves with error bands 
are plotted using the best fit parameters. It is seen that the function forms mentioned above 
also describe the data very well. The fitted results are shown in Table~\ref{table:fit} and can be compared with the $\gamma_5$-case directly. On both ensembles, it is clear that the results of the two cases are compatible with each other within errors. 

For each ensemble, the final results are obtained by averaging the results of the $\gamma_5$ and $\gamma_5\gamma_4$ case weighted by their inverse error square. To be specific, for a quantity $A$, its averaged value and error are calculated through
\begin{equation}
A_{\texttt{avg}} = \sum_i \omega_i A_i,~~~~
A_{\texttt{err}} = \sqrt{\sum_{i,j} \omega_i\omega_j \sigma_i\sigma_j},
\end{equation}
with $\omega_i = \sigma_i^{-2}/\sum_i \sigma_i^{-2}$, where $\sigma_i$ is the standard error of observable $A_i$. The averaged results are also shown in the rows labelled by \texttt{avg.} in Table~\ref{table:fit}. Finally, we get the following mixing angles $\theta_1$ and the mixing energies $x_1$ on ensemble \texttt{I} and \texttt{II} 
\begin{eqnarray}
\theta_1&=&6.6(9)^\circ,~~~x_1=48(9)~\mathrm{MeV}~~~(\mathrm{Ensemble~~I})\nonumber\\
\theta_1&=&4.3(4)^\circ,~~~x_1=49(6)~\mathrm{MeV}~~~(\mathrm{Ensemble~~II}).
\end{eqnarray}
Since the mass $m_{\eta_1}$ on ensemble \texttt{II} is close to the experimental value of $m_{\eta_c}$, we use Eq.~(\ref{eq:mix-angle}) to estimate the 
mass shift from $m_{(c\bar{c})_1}$ due to the mixing as 
\begin{eqnarray}\label{eq:shift_lat}
\Delta m^{lat}_{\eta_c}&\approx& \frac{x_1^2}{m_{\eta_1}-m_{g_1}} \approx 3.4(9)~\mathrm{MeV}.
\end{eqnarray}

\section{Discussion}\label{sec:discussion}
Till now, our major conclusion is that there does exist the mixing between the ground state pseudoscalar glueball and the ground state pseudoscalar charmonium. In this section, we will discuss the possible phenomenological implications of this kind of mixing. The prerequisite of these discussions is the identification of the possible 
pseudoscalar glueball candidate in experiments. The BESIII collaboration has observed a likely pseudoscalar structure $X(2370)$ in the processes $J/\psi\to \gamma \eta' \pi\pi$~\cite{BESIII:2010gmv} and $J/\psi \to\gamma \eta'K\bar{K}$~\cite{BESIII:2019wkp}.
The mass of $X(2370)$ is consistent with the prediction of the pseudoscalar glueball mass from lattice QCD studies. On the other hand, the branching fractions are measured to be $\mathrm{Br}(J/\psi\to \gamma X\to \gamma \eta'K^+K^-)=(1.79\pm0.23(\mathrm{stat.})\pm 0.65(\mathrm{syst.}))\times 10^{-5}$ and $\mathrm{Br}(J/\psi\to \gamma X\to \gamma \eta' K_SK_S)=(1.18\pm0.32(\mathrm{stat.}))\pm 0.39(\mathrm{syst.}))\times 10^{-5}$~\cite{BESIII:2019wkp}, which are also compatible with the lattice prediction of the branching ratio of $J/\psi$ radiatively decaying into a pseudoscalar glueball, namely
$\mathrm{Br}(J/\psi\to \gamma G)=2.31(80)\times 10^{-4}$~\cite{Gui:2019dtm}. With these observations, we tentatively consider $X(2370)$ as the pseudoscalar glueball candidate in the following discussions.

First, we consider the mass shift of the pseudoscalar charmonium due to the mixing. The non-zero mixing angle $\theta_1$ and mixing energy $x_1$ imply that the mixing can shift the masses of the unmixed pseudoscalar charmonium upward. Since $x_1$ reflects the dynamics of the $c\bar{c}$-glueball mixing, it is expected that $x_1$ is insensitive to the small variances of the masses of the pseudoscalar glueball and charmonium. Thus, to the lowest order of the perturbation theory (see Eq.~(\ref{eq:mix-angle})), the mixing angle and the mass shift of the ground state pseudoscalar charmonium can be estimated as 
\begin{eqnarray}\label{eq:shift_exp}
\sin \theta^{exp} &\approx& \frac{x_1}{m_{\eta_c}-m_{X(2370)}} \approx 0.080(10)\nonumber\\
\Delta m^{exp}_{\eta_c}&\approx& \frac{x_1^2}{m_{\eta_c}-m_{X(2370)}}\approx 3.9(9)~\mathrm{MeV},
\end{eqnarray}
where the corresponding mixing angle is $\theta^{exp}\approx4.6(6)^\circ$. These results are relevant to the charmonium hyperfine splitting $\Delta_\mathrm{HFS}=m_{J/\psi}-m_{\eta_c}$, which is usually used as a good quantity to calibrate the systematic uncertainties of lattice QCD calculations in charm physics. The PDG2020 result~\cite{Zyla:2020zbs} gives $\Delta_\mathrm{HFS}=113.0(4)$ MeV. The latest lattice calculation carried out by the HPQCD collaboration finds $\Delta_\mathrm{HFS}=120.3(1.1)$ MeV at the physical point after considering the quenched QED effects~\cite{Hatton:2020qhk}. Obviously, this result, with a much smaller error, still deviates the experimental value by $+7.3(1.2)$ MeV. The uncontrolled systematic uncertainties of this calculation are the charm quark annihilation effects and the possible mixing between pseudoscalar glueball and the pseudoscalar charmonium. As far as the charm annihilation effects are concerned, previous lattice studies show that they contribute little to the $J/\psi$ mass while move the $\eta_c$ mass upward by roughly 2 MeV~\cite{Levkova:2010ft}. We have also investigated this effects using the same ensembles in this work and obtained the mass shift of $\eta_c$ due to the charm annihilation effects is $+3.7(5)$ MeV~\cite{Zhang:2021xrs}. We are not sure whether this mass shift is theoretically equivalent to $\Delta m_{\eta_c}$ in Eq.~(\ref{eq:shift_lat}) or they can be combined together to give the total mass shift of $\eta_c$. Anyway, these corrections to the $\eta_c$ mass are in the right direction. On the other hand, the effect of light sea quarks, which are not considered in this work, may push the $\eta_c$ mass upward further. 

Secondly, the $c\bar{c}$-glueball mixing can contribute substantially to the total width of $\eta_c$. It is known that $\eta_c$ decays predominantly into light hadrons such that the total width of $\eta_c$ can be approximated as $\Gamma_{\eta_c}\approx \Gamma(\eta_c\to \mathrm{LH})$ where LH stands for all the light hadron final states. 
Although a direct derivation of $\Gamma(\eta_c\to \mathrm{LH})$ cannot be achieved from lattice QCD in the present era, we can discuss the contribution of the $c\bar{c}$-glueball mixing to $\Gamma_{\eta_c}$ by the following logic. To the leading order of perturbative QCD, the processes $\eta_c\to \mathrm{LH}$ can be viewed as that $\eta_c$ decays into two gluons first and then the two gluons are hadronized into light hadrons. Thus the decay width $\Gamma(\eta_c\to \mathrm{LH})$ can be expressed as 
\begin{equation}
    \Gamma_{\eta_c}\approx \Gamma(\eta_c\to \mathrm{LH})\approx \Gamma(\eta_c\to g g)=\frac{1}{2}\frac{1}{16\pi}\frac{1}{m_{\eta_c}}|\mathcal{M}(\eta_c\to g g)|^2
\end{equation}
where the additional factor is due to the identical two final state gluons. These arguments also apply to the hadronic decays of the pseudoscalar glueball (denoted by $|G\rangle$) and charmonium (denoted by $|(c\bar{c})\rangle$). Therefore we obtain the following relation 

\begin{equation}\label{eq:decayratio}
\frac{|\mathcal{M}(G\to g g)|}{|\mathcal{M}(c\bar{c}\to g g)|}\approx \left(\frac{m_G\Gamma_G}{m_{c\bar{c}}\Gamma_{c\bar{c}}}\right)^{1/2},
\end{equation} 
where $(m_G, \Gamma_G)$,$(m_{c\bar{c}},\Gamma_{c\bar{c}})$ are the mass and width of $|G\rangle$ and $|(c\bar{c})\rangle$, respectively.
If $\eta_c$ is an admixture of $|G\rangle$ and $|(c\bar{c})\rangle$, i.e. $|\eta_c\rangle = \cos\theta|(c\bar{c})\rangle + \sin\theta|G\rangle$ (see Eq.~(\ref{eq:mix})), by using Eq.~(\ref{eq:decayratio}) the ratio of $\Gamma_{\eta_c}$ to $\Gamma_{c\bar{c}}$ is expressed as 
\begin{eqnarray}\label{eq:widthratio}
\frac{\Gamma_{\eta_c}}{\Gamma_{c\bar{c}}}\approx \frac{|\mathcal{M}(\eta_c\to g g)|^2 m_{c\bar{c}}}{|\mathcal{M}(c\bar{c}\to g g)|^2 m_{\eta_c}}
&\approx& \left|\cos \theta +\sin\theta \frac{\left|\mathcal{M}(G\to g g)\right|}{\left|\mathcal{M}(c\bar{c}\to g g)\right|}\right|^2\nonumber\\
&\approx& 1+2\sin\theta \left(\frac{m_G\Gamma_G}{m_{\eta_c}\Gamma_{\eta_c}}\right)^{1/2}\left(\frac{\Gamma_{\eta_c}}{\Gamma_{c\bar{c}}}\right)^{1/2},
\end{eqnarray}
where we use $m_{c\bar{c}}\approx m_{\eta_c}$, $\cos\theta\approx1$. With the assumption that $X(2370)$ is predominantly a pseudoscalar glueball, 
and if we take $\Gamma_G\approx\Gamma_{X(2370)}\approx 100$ MeV and use the PDG value $\Gamma_{\eta_c}= 32.0(7)$ MeV, then by solving Eq.~(\ref{eq:widthratio}) we get $\frac{\Gamma_{\eta_c}}{\Gamma_{c\bar{c}}}=1.29(4)$, which implies $\Gamma_{c\bar{c}}\approx 24.8(9)$ MeV. Finally, the contribution  of $c\bar{c}$-glueball mixing to $\Gamma_{\eta_c}$ is estimated as 
\begin{equation}
    \delta \Gamma_{c\bar{c}}\equiv\Gamma_{\eta_c}-\Gamma_{c\bar{c}}\approx 7.2(8)~~\mathrm{MeV}.
\end{equation}

The decays of $c\bar{c}$ pseudoscalar meson into hadrons can be viewed as that the $c\bar{c}$ first decays into two gluons and then the two gluons are hadronized into light hadrons.  In this sense, one can take the approximation $\Gamma_{c\bar{c}}\approx \Gamma(c\bar{c}\to gg)$. On the other hand, the radiative decay $\eta_c\to \gamma\gamma$ is dominated by $c\bar{c}\to \gamma\gamma$. According to the running of the strong coupling constant $\alpha_s(\mu)$, at $\mu\approx m_c\approx 1.5$ GeV, $\alpha_s$ takes the value in the range $0.3<\alpha_s<0.35$. If one takes $\alpha=1/134$ at the charm quark mass scale, to the leading order QCD correction~\cite{Tsai:2011dp,Kwong:1987ak} one has 
\begin{equation}\label{eq:ratio}
\frac{\Gamma(c\bar{c}\to \gamma\gamma)}{\Gamma(c\bar{c}\to gg)}\approx \frac{8}{9}\frac{\alpha^2}{\alpha_s^2}\frac{1-3.4\alpha_s/\pi}{1+4.8\alpha_s/\pi}\approx (1.6\sim2.5)\times 10^{-4}
\end{equation}
Experimentally, the PDG result of $\mathrm{Br}(\eta_c\to \gamma\gamma)=(1.61\pm0.12)\times 10^{-4}$~\cite{Zyla:2020zbs}. Considering the ratio $\Gamma_{\eta_c}/\Gamma_{c\bar{c}}=1.29(4)$, the experimental value implies ${\Gamma(c\bar{c}\to \gamma\gamma)}/{\Gamma(c\bar{c}\to gg)}\approx (2.07\pm 0.17)\times 10^{-4}$, which falls into the range of Eq.~(\ref{eq:ratio}). Note that the NRQCD analysis with next-next-leading order QCD corrections predicts the branching fraction $\mathrm{Br}(\eta_c\to\gamma\gamma)\sim (2.3-2.9)\times 10^{-4}$ if $\eta_c$ is taken as a pure $c\bar{c}$ state~\cite{Brambilla:2018tyu}, which also requires a smaller total width of the ground state pseusoscalar charmonium when comparing with PDG value. Anyway, the above discussions are just tentative because of the assumption that $X(2370)$ is predominantly a pseudoscalar glueball. The existence and the status of $X(2370)$ need to be clarified by future experiments. 

As for the decay width of $J/\psi\to \gamma \eta_c$, however, the tension between the experiments and the theoretical predictions cannot be alleviated by the $c\bar{c}$-glueball mixing. PDG gives the world average value $\mathrm{Br}(J/\psi\to\gamma\eta_c)=(1.7\pm 0.4)\times 10^{-2}$ ~\cite{Zyla:2020zbs}, which corresponds to the partial decay width $\Gamma(J/\psi\to\gamma\eta_c)=1.6\pm 0.4$ keV. Theoretically, the effective field theories and non-relativistic potential models predict the partial width to be 1.5-2.9 keV~\cite{Khodjamirian:1983gd,Godfrey:1985xj,Barnes:2005pb,Brambilla:2005zw,Brambilla:2010ey,Pineda:2013lta}. The result from the NRQCD effective field theory predicts the branching fraction to be $(1.5\pm 1.0)$ keV~\cite{Brambilla:2005zw}, which is compatible with the experimental value but with a quite large error. The predictions of most of lattice QCD calculations, both quenched and full-QCD ones~\cite{Dudek:2006ej,Dudek:2009kk,Chen:2011kpa,Becirevic:2012dc,Donald:2012ga,Gui:2019dtm}, are around 2.4-2.9 keV, which have discrepancies from the PDG value, but are in agreement with the KEDR experimental result $\Gamma(J/\psi\to\gamma\eta_c)=2.98\pm 0.18^{+0.15}_{-0.33}$ keV~\cite{Anashin:2014wva}. As addressed before, since the radiative production rate of the pseudoscalar glueball in the $J/\psi$ decays is two orders of magnitude smaller than that of the pseudoscalar charmonium~\cite{Gui:2019dtm}, and $\eta_c$ has a very small fraction of the pseudoscalar glueball, the mixing cannot change the partial width of $J/\psi\to \gamma\eta_c$. Hopefully, the contraversial situation on the decay width of $J/\psi\to \gamma \eta_c$ can be resolved by the future study of the BESIII collaboration using its large $J/\psi$ event sample.  

\section{Summary}\label{sec:summary}
We generate large gauge ensembles with $N_f=2$ degenerate charm quarks on anisotropic lattices, such that the theoretical framework is unitary for charm quarks. The annihilation diagrams of charm quark are tackled through the distillation method. By calculating the correlation functions of the pseudoscalar quark bilinear operators and the pseudoscalar glueball operator, the mixing energy $x= 49(6)$ MeV and the mixing angle $\theta= 4.3(4)^\circ$ have been obtained for the first time through lattice QCD calculations. 

The nonzero mixing energy and the mixing angle help to understand the properties of the $\eta_c$ meson. If $X(2370)$ observed by BESIII can be taken as predominantly a pseudoscalar glueball, then the $c\bar{c}$-glueball mixing can result in a positive mass shift approximately $3.9(9)$ MeV of the ground state pseudoscalar charmonium, which serves to understand the discrepancy of lattice and the experimental results of the $1S$ hyperfine splitting of charmonia. In the mean time, the mixing implies that the total width of the pseudoscalar charmonium can be increased by 7.2(8) MeV, which can explain to some extent the relatively large width of $\eta_c$ in comparison with the theoretical expectations for a pure $c\bar{c}$ state. Furthermore, the branching fraction of $\eta_c\to \gamma\gamma$ can be understood in this $c\bar{c}$-glueball framework. It should be notified that, even though the assumption that $X(2370)$ is predominantly a pseudoscalar glueball seems compatible with the discussion in this work, its justification should be clarified by future experimental and theoretical investigations. At last, the seemingly discrepancy of the theoretical predictions and the experimental results of the partial width of $J/\psi\to\gamma\eta_c$ cannot be alleviated by the $c\bar{c}$-glueball mixing picture, which demands future sophisticated experimental studies. The BESIII collaboration may take this mission by the help of its largest $J/\psi$ event ensemble in the world.   
\vspace{0.5cm}
\section*{Acknowledgements} 
We thank Prof. Q. Zhao of IHEP for the inspiring discussions. This work is supported by the Strategic Priority Research Program of Chinese Academy of Sciences (No. XDB34030302), the National Key Research and Development Program of China (No. 2020YFA0406400) and the National Natural Science Foundation of China (NNSFC) under Grants No.11935017, No. 11575196, No.11775229, No.12075253, No.12070131001 (CRC 110 by DFG and NNSFC), No.12175063. Y. Chen also acknowledges the support of the CAS Center for Excellence in Particle Physics (CCEPP). L.C. Gui is supported by Natural Science Foundation of Hunan Province under Grants No.2020JJ5343, No.20A310. The Chroma software system~\cite{Edwards:2004sx} and QUDA library~\cite{Clark:2009wm,Babich:2011np} are acknowledged. The computations were performed on the CAS Xiandao-1 computing environment, the HPC clusters at Institute of High Energy Physics (Beijing) and China Spallation Neutron Source (Dongguan), and the GPU cluster at Hunan Normal University.

\bibliographystyle{elsarticle-num}
\bibliography{ref}

\section*{Appendix}
\begin{appendices}

\section{Glueball operator construction}\label{appendix:oper}
\begin{figure*}[htbp]
    \centering
	\includegraphics[width=0.5\linewidth]{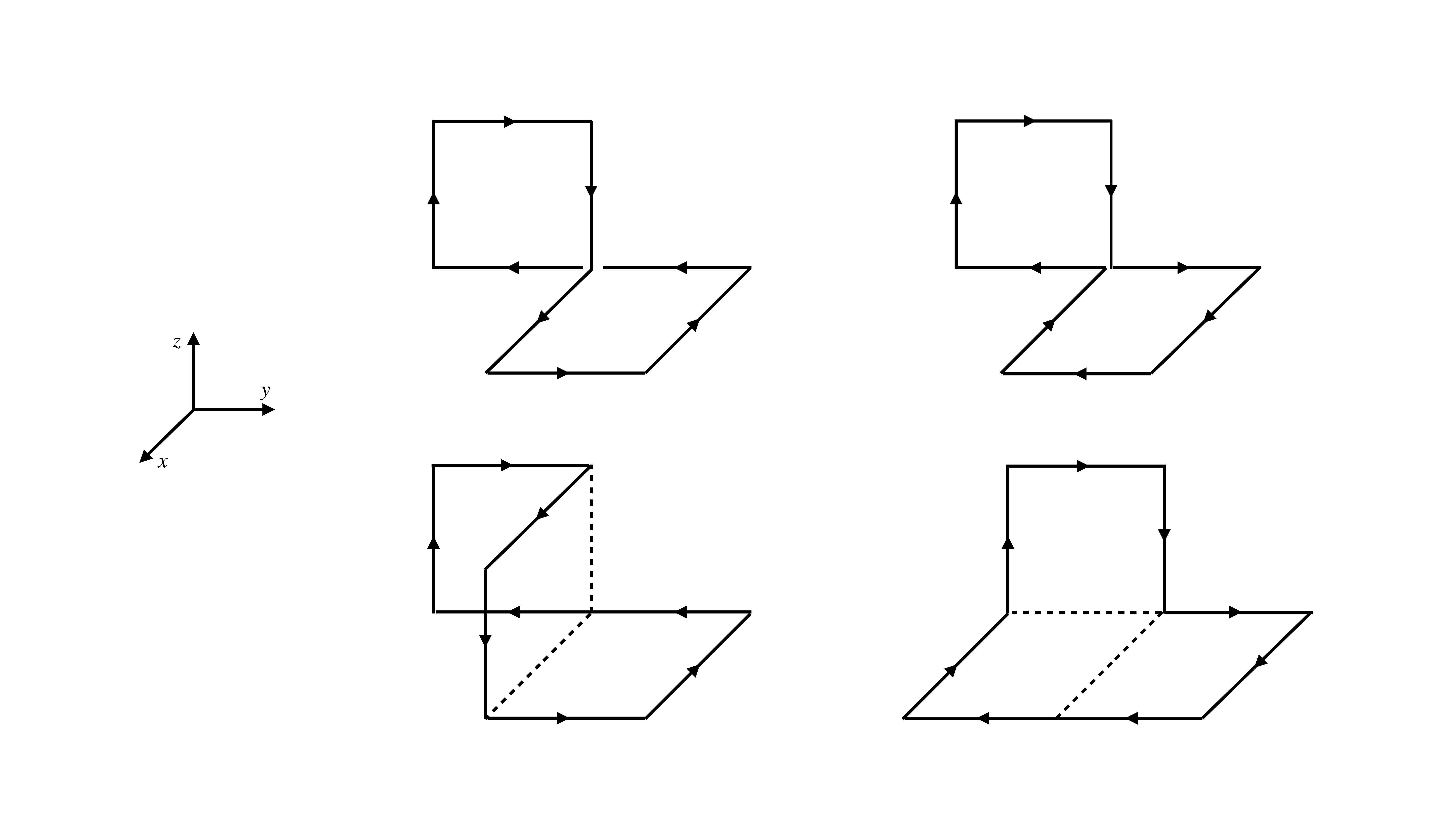}\hspace{1cm}
	\caption{\label{fig:O_G}Wilson loop prototypes used to construct the pseudoscalar glueball operator~\cite{Morningstar:1999rf,Chen:2005mg}.}
\end{figure*}

Based on the four prototypes of Wilson loops shown in Fig.~\ref{fig:O_G}, we build $A_1^{-+}$ operators for the pseudoscalar glueball operators 
with $A_1$ being an irreducible representation of spatial symmetry group, namely, the octahedral group $O$. 
We adopt six different schemes to smear gauge links, which are different combinations of single-sink smearing and double link smearing~\cite{Morningstar:1999rf,Chen:2005mg}. Let $W_\alpha(\mathbf{x},t)$ be one prototype of Wilson loop under a specific smearing scheme, then the $A_1^{-+}$ operator in the rest frame of a glueball can be obtained by 
\begin{equation}
    \phi_\alpha (t)=\sum\limits_{\mathbf{x}}\sum\limits_{R\in O} c_R^{A_1}\left[R\circ W_\alpha(\mathbf{x},t)-\mathcal{P}R\circ W_\alpha(\mathbf{x},t)\mathcal{P}^{-1}\right]
\end{equation}
where $R\circ W_\alpha$ refers to a differently oriented Wilson loops after one of the 24 elements of $O$ ($R$) operated on $W_\alpha$, $\mathcal{P}$ is spatial reflection operation and $C_R^{A_1}$ are the combinational coefficients for the $A_1$ representation. Thus we obtain a $A_1^{-+}$ operator set $\{\phi_\alpha(t),\alpha=1,2,\ldots, 24\}$ based on the four prototypes and six smearing schemes. 
We then use the well established variational method to get an optimized operator $O_G$ that couples most to the ground state glueball by solving the generalized eigenvalue problem. 

\section{Large $t$ behaviour of $C_{GG}(t)$ and $C_{CC}(t)$}\label{appendix:comparison}
\begin{figure*}[htbp]
    \centering
	\includegraphics[width=0.45\linewidth]{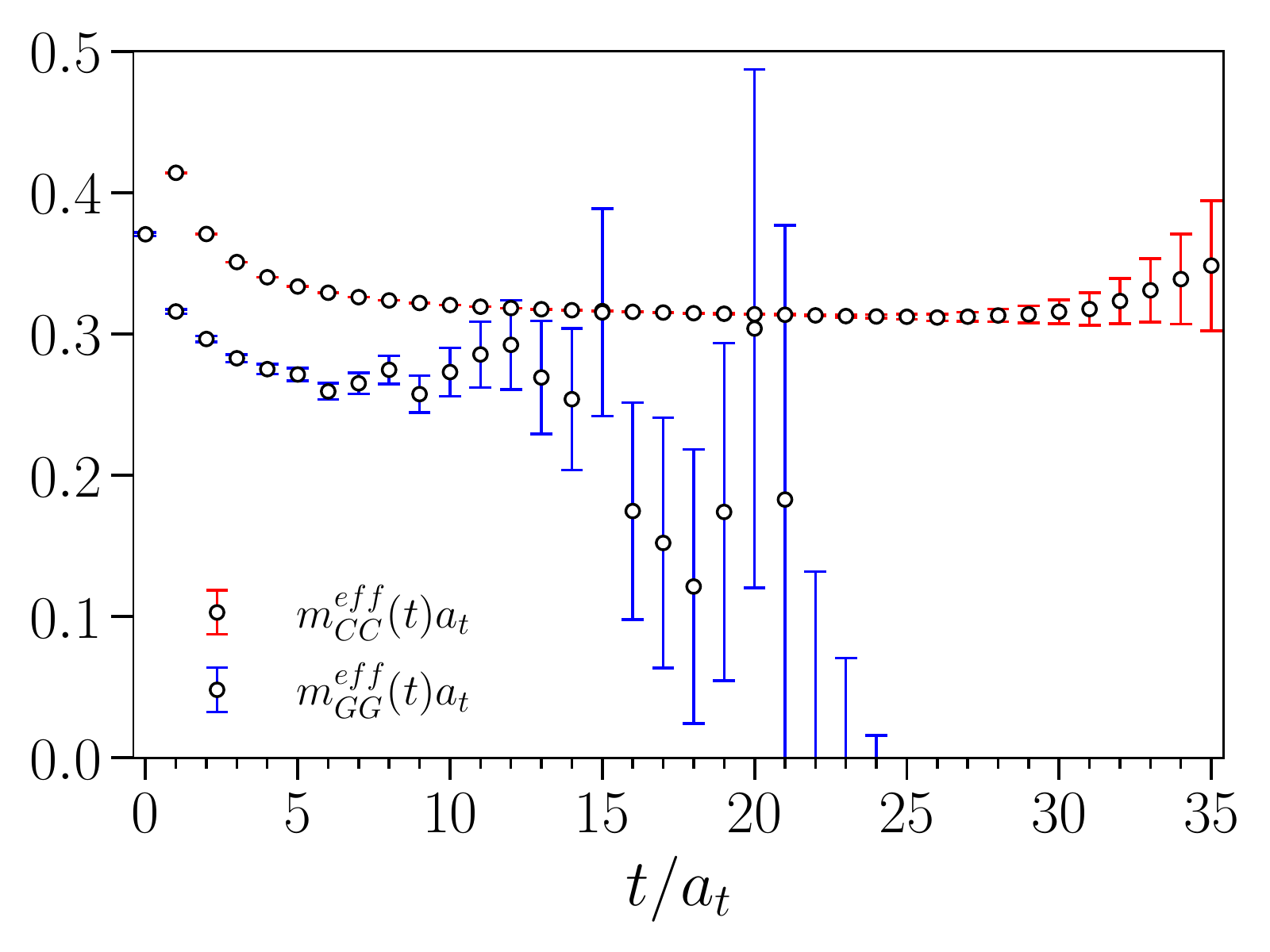}\hspace{1cm}
	\includegraphics[width=0.45\linewidth]{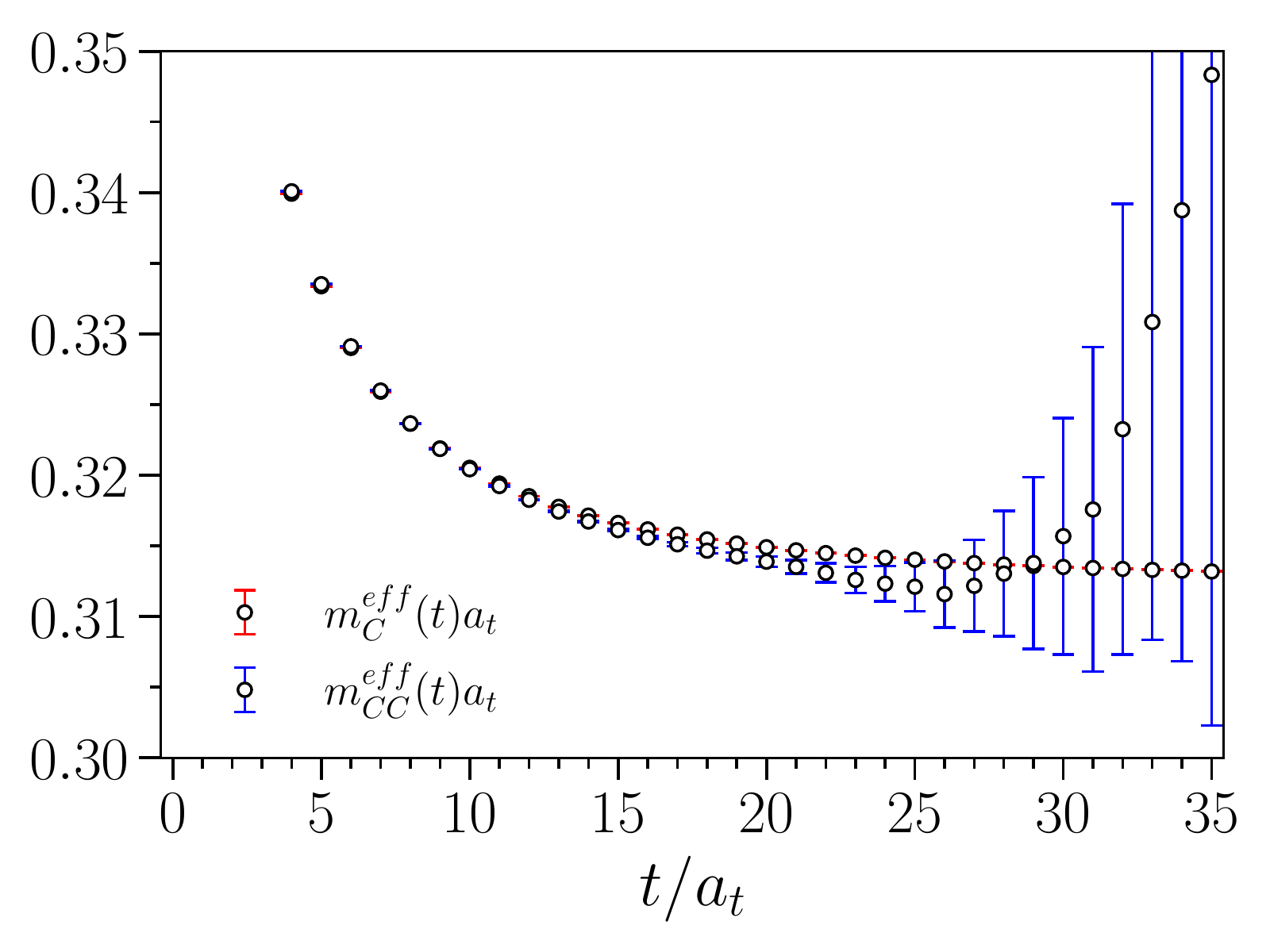}
	\caption{\label{fig:effemgamma5_CC_GG}(left): Effective mass from two point functions 
		$C_{CC}(t)$, $C_{GG}(t)$ for operator with $\Gamma=\gamma_5$ on ensemble \texttt{II} shown in same plot for comparison.
		(right): Effective mass $m^\mathrm{eff}_{C}(t)$ from connected part of $O_\Gamma$'s correlation function $C(t)$ and $m^\mathrm{eff}_{CC}(t)$ from full correlation function $C(t)+2D(t)$ including disconnected diagram with $\Gamma=\gamma_5$ on ensemble \texttt{II}. }
\end{figure*}

The spectrum of $C_{GG}(t)$ and $C_{CC}(t)$ should be the same, as reflected by Eq.~(\ref{eq:gg-gamma5}). It is expected $C_{GG}(t)$ and $C_{CC}(t)$ are dominated by the contribution from the lowest state $|g_1\rangle$ at very large $t$ (when $T\to \infty$) such that their effective masses $m^\mathrm{eff}_{GG}(t)$ and $m^\mathrm{eff}_{CC}(t)$ should merge into $m_{g_1}a_t$ at very large $t$. The left panel of Fig.~\ref{fig:effemgamma5_CC_GG} shows $m^\mathrm{eff}_{GG}(t)$ and $m^\mathrm{eff}_{CC}(t)$ for the $\Gamma=\gamma_5$ case on ensemble \texttt{II}. We do not observed a clear tendency that they will merge together at $m_{g_1}a_t$ in the available time range. This implies that $\sin^2 \theta_1$ should be very small, as confirmed by the fitted result $\sin^2\theta_1\sim 0.006$. 

On the other hand, the error of $m^\mathrm{eff}_{CC}(t)$ is quite small when $t/a_t<20$ but grows rapidly beyond $t/a_t>25$. Actually, the error of $C_{CC}(t)$ comes mainly 
from the error of the disconnected diagram contribution.  Since the disconnected part falls off more slowly than the connected part, the contribution and the error of the former become more pronounced when $t$ increases. In the right panel of Fig.~\ref{fig:effemgamma5_CC_GG}, we compare $m^\mathrm{eff}_{CC}(t)$ with the effective mass $m^\mathrm{eff}_{C}(t)$ of the connected part for the $\Gamma=\gamma_5$ case on ensemble \texttt{II}. It is seen that $m^\mathrm{eff}_{CC}(t)$ deviates from $m^\mathrm{eff}_{C}(t)$ gradually beyond $t/a_t\sim 15$ where the mixing effect begins to show up.


\end{appendices}

\end{document}